\documentclass[aps,prb,floatfix,twocolumn,10pt,superscriptaddress,longbibliography]{revtex4-1}

\usepackage{graphicx}
\usepackage{dcolumn}
\usepackage{soul}
\usepackage{color}
\usepackage[svgnames]{xcolor}
\usepackage{amssymb}
\usepackage{float}
\usepackage{natmove}
\usepackage{multirow}
\usepackage{setspace}
\usepackage{array}
\usepackage{booktabs}
\usepackage{rotating}
\usepackage[colorlinks=true,linkcolor=blue,citecolor=blue,urlcolor=blue]{hyperref}
\usepackage{}
\usepackage{physics}
\usepackage{mathrsfs}
\usepackage{amssymb,amsfonts,dsfont}
\usepackage{bbold}
\usepackage{times} 
\usepackage{bm} 
\usepackage{amsmath} 
\usepackage{accents} 
\usepackage{epic}
\usepackage{wrapfig}
\usepackage{ulem}
\usepackage{tabularx}
\usepackage{colortbl}
\makeatletter

    \def\CT@@do@color{%
      \global\let\CT@do@color\relax
            \@tempdima\wd\z@
            \advance\@tempdima\@tempdimb
            \advance\@tempdima\@tempdimc
    \advance\@tempdimb\tabcolsep
    \advance\@tempdimc\tabcolsep
    \advance\@tempdima2\tabcolsep
            \kern-\@tempdimb
            \leaders\vrule
                    \hskip\@tempdima\@plus  1fill
            \kern-\@tempdimc
            \hskip-\wd\z@ \@plus -1fill }
    \makeatother
    
\newcommand{\editor}[2]{%
  \expandafter\newcommand\csname #1note\endcsname[1]{%
    \textcolor{#2}{(\textbf{#1:} ##1)}}%
  \expandafter\newcommand\csname #1\endcsname[1]{%
    \textcolor{#2}{##1}}%
  \expandafter\newcommand\csname #1cancel\endcsname[1]{%
    \textcolor{#2}{\sout{##1}}}%
  \expandafter\newcommand\csname #1change\endcsname[2]{%
    \textcolor{#2}{\sout{##1} ##2}}%
  \newenvironment{#1text}{\color{#2}}{\color{black}}
}
\definecolor{Blu}{rgb}{0.00,0.00,1.00}
\definecolor{Red}{rgb}{1.00,0.00,0.00}
\definecolor{Orange}{rgb}{0.95,0.46,0.17}
\editor{DV}{blue}
\editor{MR}{Red}
\editor{EM}{Magenta}
\editor{GS}{Orange}

\begin{document}
\title{Anomalous screening in narrow-gap carbon nanotubes}
\author{Giacomo Sesti}
\affiliation{FIM, Universit\`a degli Studi di Modena e Reggio Emilia, Via Campi 213a, 41125 Modena, Italy}
\affiliation{CNR-NANO, Via Campi 213a, 41125 Modena, Italy}
\author{Daniele Varsano}
\affiliation{CNR-NANO, Via Campi 213a, 41125 Modena, Italy}
\author{Elisa Molinari}
\affiliation{FIM, Universit\`a degli Studi di Modena e Reggio Emilia, Via Campi 213a, 41125 Modena, Italy}
\affiliation{CNR-NANO, Via Campi 213a, 41125 Modena, Italy}
\author{Massimo Rontani}
\affiliation{CNR-NANO, Via Campi 213a, 41125 Modena, Italy}

\begin{abstract}
The screening of Coulomb interaction controls many-body physics in carbon nanotubes, as it tunes the range and strength of the force that acts on charge carriers and binds electron-hole pairs into excitons. In doped tubes, the effective Coulomb interaction drives the competition between Luttinger liquid and Wigner crystal, whereas in undoped narrow-gap tubes it dictates the Mott or excitonic nature of the correlated insulator observed at low temperature.  
Here, by computing the dielectric function of selected narrow- and zero-gap tubes from first principles, we show that the standard effective-mass model of screening systematically underestimates the interaction strength at long wavelength, hence missing the binding of low-energy excitons. The reason is that the model critically lacks the full three-dimensional topology of the tube, being adapted from graphene theory. As ab inito calculations are limited to small tubes, we develop a two-band model dielectric function based on the plane-wave expansion of Bloch states and the exact truncated Coulomb cutoff technique. We demonstrate that our---computationally cheap---approach provides the correct screening for narrow-gap tubes of any size and chirality. A striking result is that the screened interaction remains long-ranged even in gapless tubes, as an effect of the microscopic local fields generated by the electrons moving on the curved tube surface. As an application, we show that the effective electron-electron force that is felt at distances relevant to quantum transport experiments is super Coulombic.         
\end{abstract}

\maketitle

\section{Introduction}

Carbon nanotubes (NTs) are ideal materials to study strongly interacting electrons in one dimension.\cite{Dresselhaus1998,Charlier2007,McEuen2010,Laird2015} Due to the reduced dimensionality, Coulomb interaction remains unscreened at long wavelength whereas its strength reaches extreme values in suspended, undoped tubes, as the electric field lines spill over into the vacuum.
Excitons---electron-hole pairs bound by Coulomb attraction---exhibit huge binding energies\cite{Ando1997,Maultzsch2005,Wang2005,spataru2004excitonic,capaz2006diameter} that remain finite even in metallic NTs.\cite{Wang2007,deslippe2007bound} Ultraclean, narrow-gap NTs may be charged in a controlled and reversible way by means of Coulomb blockade, hosting fascinating many-body states of matter. These include the Luttinger liquid,\cite{Giamarchi2004,Balents1997,Kane1997b,Egger1997,Krotov1997,Bockrath1999,Postma2001,Deshpande2010} the Wigner crystal,\cite{Deshpande2008,Secchi2009,Deshpande2010,Pecker2013,Ilanit2019,Lotfizadeh2019,Ziani2021} and a correlated insulator understood as either a Mott\cite{Deshpande2009} or an excitonic phase\cite{varsano2017carbon}
(but Peierls insulators,\cite{Bohnen2004,Connetable2005,dumont2010peierls} topological phases,\cite{Efroni2017} and hybrid scenarioes\cite{Chen2008,Hellgren2018,Okamoto2018} were proposed as well).   

The focus of this work is the quantitative assessment of the screened  Coulomb interaction in narrow-gap NTs, as the effective range and strength of the electron-electron force rule the nature of the correlated phases observed in quantum transport experiments. In slightly doped tubes, the long range part of Coulomb interaction stabilizes the Wigner localization of electrons (or holes), its strength tuning the charge modulation associated with crystal-like order.\cite{Schulz1993,secchi2010wigner,Secchi2012,Wang2012b} As screening controls the force range, one may melt the crystal e.g. by increasing doping or changing the dielectric environment, leaving room to Luttinger liquid. The elementary excitations of the liquid are plasmons, whose velocities depend on the residual, short-range interaction.\cite{Giamarchi2004}    

Undoped NTs are always insulating,\cite{Deshpande2009,Deshpande2010,McEuen2010,Laird2015,Senger2018,Island2018} including the armchair kind, which band theory predicts to be metallic and protected against gap-opening perturbations.\cite{Charlier2007} The contribution to the gap that is not accounted for by independent-electron models is thought to have a many-body origin, whose features---again---critically depend on the range of electron-electron interaction. One possible conventional scenario is the Mott insulator,\cite{Deshpande2009} whose gap originates from the short-range part of Coulomb interaction. Its theory---a strong-coupling version of the Luttinger liquid---assumes the long-wavelength Coulomb force to be cutoff by nearby electrostatic gates in the experimental setup.\cite{Giamarchi2004,Balents1997,Kane1997b,Krotov1997,Nersesyan2003} A second possibility, recently proposed by some of us,\cite{varsano2017carbon} is that the residual gap is due to the long-range part of Coulomb interaction, which binds electrons and holes into excitons. If the Bohr radius is smaller than the cutoff length, then excitons condense at thermodynamic equilibrium---in the absence of optical excitation---giving rise to the "excitonic insulator" phase predicted in the sixties.\cite{Sherrington1968} 
Pivotal to this prediction is the result that the screened Coulomb interaction in momentum space, $W(q)$, exhibits a seemingly singular-like profile for $q\rightarrow 0$ in gapless tubes, as found from first principles [see inset of Fig. 2c of Ref.~\onlinecite{varsano2017carbon} and Fig.~\ref{potarmchair-2} for (3,3) and (5,5) armchair tubes, respectively]. This finding, which motivates the present work, is surprising, as the simple Thomas-Fermi model of a one-dimensional metal predicts $W(q)$ to be almost constant for $q\approx 0$.

Screening in NTs shows a complex behaviour due to its non local character, the interband electronic polarization being effective at intermediate ranges only.\cite{leonard2002dielectric} As a result, in undoped NTs the electron-hole interaction, once projected onto the lowest conduction and highest valence band, is enhanced with respect to the bare force, for carrier separation larger than the NT radius.\cite{Deslippe2009} Whereas the precise assessment of the dressed interaction requires the accurate calculation of the dielectric matrix, the computation from first principles is limited to NTs having a small unit cell, due to the heavy computational load. Therefore, a model dielectric function based on the effective-mass (EM) approximation is widely used, either in its original form by Ando\cite{Ando1997,tomio2012interwall} or in simpler versions.\cite{Deslippe2009,thakur2017dynamical} These models are able to reproduce the main nonlocal features of screening but rule out the excitonic instability.\cite{Ando1997}

Here we show from first principles that the EM model underestimates the strength of Coulomb interaction at long wavelength in both narrow- and zero-gap NTs---hence missing the binding of low-energy excitons. Crucially, the EM theory neglects the actual topology of the orbitals involved in the calculation of the polarization, as it takes Bloch states from graphene. To overcome this problem, we introduce a two-band, computationally cheap model of the polarization that copes with the tube-like topology of Bloch states by expanding them over a three-dimensional plane-wave basis set. Furthermore, we apply an exact cutoff technique to Coulomb potential,\cite{rozzi2006exact,ismail2006truncation} in order to avoid spurious interactions among replicas in our supercell calculation. We eventually validate our model through comparison with first-principles results for zigzag and armchair NTs of different radii. As a generic outcome, we find that the change of overlap integrals between conduction and valence Bloch states induced by tube curvature leads to a significant enhancement of the dressed Coulomb interaction at long wavelength, $W(q\sim 0)$. This effect is the signature of microscopic local fields, which are strong as electrons actually move on a cylindrical surface and not on a line. As a consequence, the effective attraction between electrons and holes remains long-ranged even in gapless tubes. 

Our findings support our previous claim of excitonic instability,\cite{varsano2017carbon} suggesting that the long-range part of Coulomb interaction rules many-body physics of NTs. 
As an application of the proposed model dielectric function, and motivated by a recent experiment by Shapir and coworkers at the Weizmann Institute of Science,\cite{shapireeinteraction}
we compute the real-space effective Coulomb force between two electrons populating the lowest conduction band. This observable has been measured, for various electron-electron separations, in ultraclean suspended NTs in a non invasive manner.\cite{shapireeinteraction} The agreement between numerical and experimental results will be shown elsewhere.
Our results confirm that the effective electron-electron interaction in NTs is of super Coulombic nature.

The paper is organized as follows: we explain the methodology in the first three sections and discuss the results in the remainder of the manuscript. In detail, we first illustrate the first-principles methodology (Sec.~\ref{sec:ab_initio}), then review the effective-mass approximation (Sec.~\ref{sec:EMwfs}), and  eventually detail the proposed two-band model of screening (Sec.~\ref{sec:ourTheo}). We report results starting from the bare electron-hole interaction, projected onto conduction and valence bands, in Sec.~\ref{sec:bare}. The most important findings concern the inverse dielectric function (Sec.~\ref{sec:die}) and the dressed electron-hole interaction (Sec.~\ref{sec:dressed}). We dedicate a whole subsection to the electron-hole interaction in gapless tubes (Sec.~\ref{sec:armchair}), which has profound implications for the instability of the many-body ground state towards exciton condensation. In Sec.~\ref{sec:jellium} we validate the three-dimensional structural model used throughout the work by using refined results for zigzag and armchair tubes as a benchmark. We finally present the calculation of the electron-electron force in real space
(Sec.~\ref{sec:super}), projected onto the lowest conduction band, as an experimentally relevant application. We draw our conclusions in Sec.~\ref{sec:conclusions}.  



\section{Calculations from first principles}\label{sec:ab_initio}

Calculations from first principles proved to be very reliable to study electronic properties of physical systems. \cite{Onida2002,gross2013density}  In this work, we use ab-initio results as a benchmark to investigate screening properties of selected carbon NTs. The systems considered are the (3,3) and (5,5) armchair NTs as well as the (9,0) and the (12,0) zigzag NTs. Our calculations from first principles are performed in two steps. In first instance, we perform density functional theory (DFT) computations of the NTs. On top of the DFT computation, we then compute the dielectric function and the screened potential. The real-space screened potential is reconstructed by performing an expansion over the reciprocal lattice basis: 
\begin{align}
\label{eq.Wtot}
W(\boldsymbol{r},\boldsymbol{r'})= & \notag  \sum_{\boldsymbol{q}} \sum_{\boldsymbol{G}} \sum_{\boldsymbol{G'}}  e^{i (\boldsymbol{G} + \boldsymbol{q}) \cdot \boldsymbol{r}}  e^{-i (\boldsymbol{G'} + \boldsymbol{q}) \cdot \boldsymbol{r'}} \epsilon^{-1}_{\boldsymbol{G},\boldsymbol{G'}}(\boldsymbol{q},0) \\ &   \times \quad v(\boldsymbol{q} + \boldsymbol{G'}),
\end{align}
where $\epsilon^{-1}_{\boldsymbol{G},\boldsymbol{G'}}(\boldsymbol{q},\omega)$ is the momentum- and frequency-dependent inverse dielectric matrix, $v(\boldsymbol{q}) =
4\pi e^2 \Omega^{-1} /q^2$ is the bare Coulomb potential, $\boldsymbol{G}$ is the reciprocal lattice vector, and $\Omega$ is the system volume. We treat the screening within the random phase approximation (RPA):
\begin{eqnarray}
\label{RPA_1}
\epsilon^{-1}_{\boldsymbol{G},\boldsymbol{G'}}(\boldsymbol{q},\omega) = \big[ \delta_{\boldsymbol{G},\boldsymbol{G'}} - \Pi_{\boldsymbol{G},\boldsymbol{G'}}(\boldsymbol{q},\omega)\, v(\boldsymbol{q} + \boldsymbol{G}) \big]^{-1},
\end{eqnarray}
with
$\Pi_{\boldsymbol{G},\boldsymbol{G'}}(\boldsymbol{q},\omega)$ being the irreducible polarisation:
\begin{align}
\label{eq.polabin}
\Pi_{\boldsymbol{G},\boldsymbol{G'}}(\boldsymbol{q},\omega) = &  \notag 2 \sum_{n,n'} \sum_{\boldsymbol{k}}  \frac{ f(E_{n,\boldsymbol{k}}) - f(E_{n',\boldsymbol{k}+\boldsymbol{q}}) }{ \omega + E_{n,\boldsymbol{k}} -E_{n',\boldsymbol{k}+\boldsymbol{q}} + i \eta} \\ & \times \quad \rho_{n,n'}^*(\boldsymbol{k},\boldsymbol{q},\boldsymbol{G'}) \,\rho_{n,n'} (\boldsymbol{k},\boldsymbol{q},\boldsymbol{G}).
\end{align}
The $f(E)$ are the occupation factors, the overlap integrals are defined as $\rho_{n,n'}(\boldsymbol{k},\boldsymbol{q},\boldsymbol{G}) = \langle  n   \boldsymbol{k}|  e^{i (\boldsymbol{G} + \boldsymbol{q}) \cdot \boldsymbol{r}} |   n' \boldsymbol{k} + \boldsymbol{q}  \rangle $, and $\eta$ is a positive infinitesimal. The indexes $n, n'$ run over the electronic bands. The energies, $E_{n,\boldsymbol{k}}$, and wavefunctions, $|   n \boldsymbol{k}  \rangle$, we employ in Eq. (\ref{eq.polabin}) are those determined by the DFT computations. As we mainly look at the long-range potential of carbon NTs, only the static polarisation $\Pi_{\boldsymbol{G},\boldsymbol{G'}}(\boldsymbol{q},0)$ is necessary in our work. \\
Density functional theory calculations were performed using the QUANTUM ESPRESSO package,\cite{QE1,QE2} where wave functions are expanded in plane waves and pseudopotentials are used to account for the electron-ion interaction. We used the local density approximation (LDA) for the exchange-correlation potential, according to the Perdew-Zunger parametrization,\cite{Perdew_PRB_1981} and norm conserving pseudopotentials. The kinetic energy cutoff to represent the Kohn-Sham wavefunction was set to 70 Ry
and an amount of vacuum of $38$ Bohr in the direction perpendicular to the nanotube axis was considered to avoid replica interactions.
The screened potential and the dielectric function were calculated using the Yambo code,\cite{Marini2009,yambo_2019} where we considered 80, 120, 300, and 600 bands in the summation of Eq.\ref{eq.polabin} for the (3,3), (5,5),  (9,0), and (12,0) NTs, respectively.
A cutoff of 4 Ry in the $\epsilon^{-1}_{G,G^\prime} $ matrix dimension was considered for all NTs. The Brillouin zone was sampled using a one dimensional grid of respectively 1973 and 205 k-points for armchair and zigzag NTs.
\section{Effective-mass theory}\label{sec:EMwfs}

\subsection{Envelope function}

\begin{figure}[t]
    \centering
    \includegraphics[width=8.6cm,trim={13cm 1cm 8cm 0.5cm},clip]{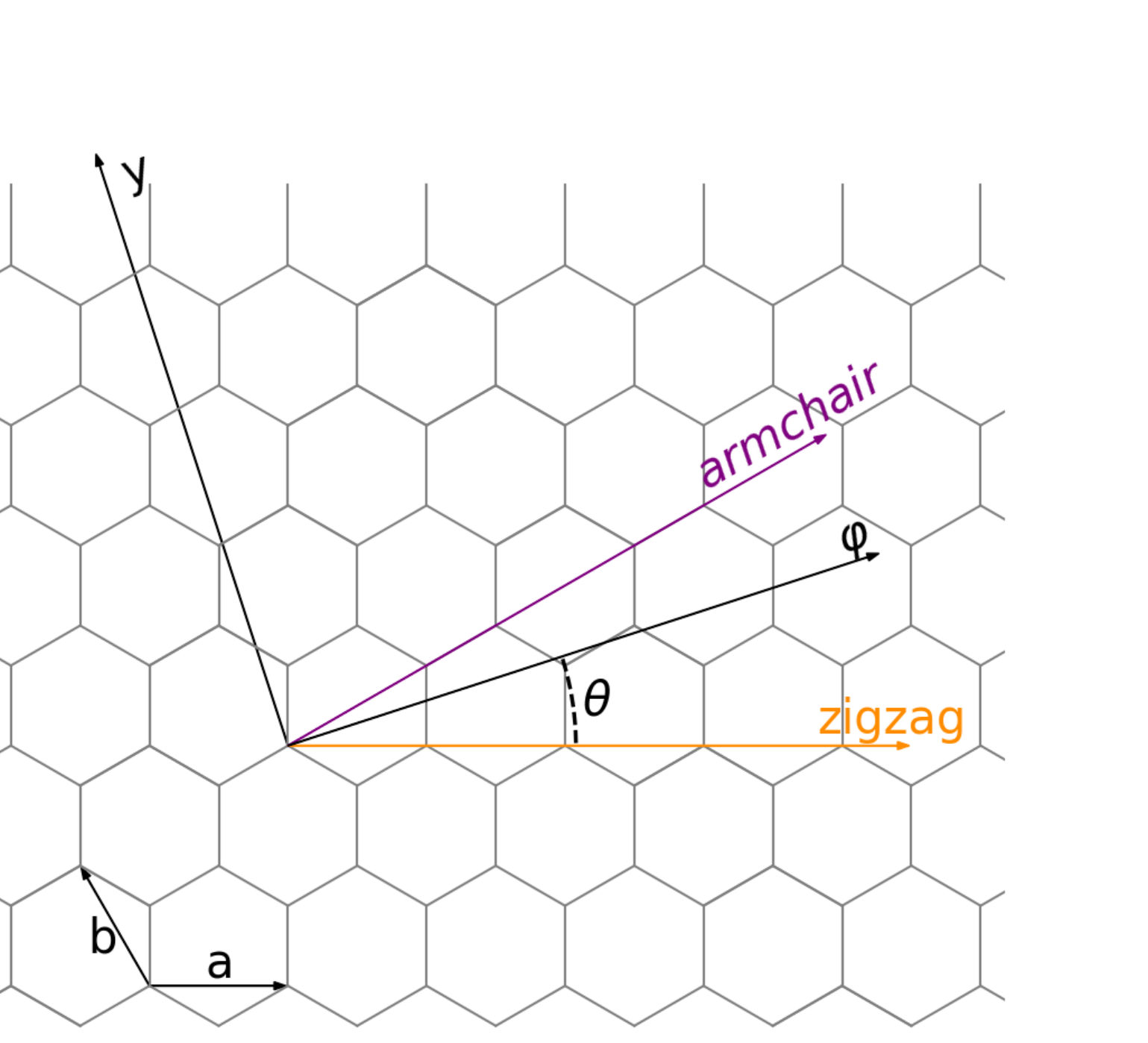}
   \caption{Pictorial illustration of the folding of a graphene sheet. Here $\varphi$ and $y$ are, respectively, the direction of folding and of the nanotube axis, whereas $a$ and $b$ are the basis vectors of the graphene lattice. The chiral angle $\theta$ spans the region between $\varphi$ and $a$. }
    \label{FIG:graphsheet}
\end{figure}

Within the effective mass (EM) and envelope function approximations, a single-wall carbon NT is treated as a rolled graphene sheet,\cite{Ajiki1993,Ando1997,Dresselhaus1998,Charlier2007} as illustrated in Fig.~\ref{FIG:graphsheet}. In the limit of large radius, the Bloch states, $\psi_{\tau} (\boldsymbol{r})$, that multiply the envelopes  coincide with the $\pi$ tight-binding states of graphene located at $\tau=$ K,K$'$ corners of the hexagonal Brillouin zone, the charge neutrality points where Dirac cones touch. For each valley $\tau$,
the NT orbital wave functions are
\begin{equation}
\label{eq.Bloch}
\Psi_{\alpha \tau   k}(\boldsymbol{r})= F^{\tau A}_{\alpha k  }(\boldsymbol{r}) \,\psi_{\tau A} (\boldsymbol{r}) + F^{\tau B}_{\alpha k  }(\boldsymbol{r})\, \psi_{\tau B} (\boldsymbol{r}),
\end{equation}
where $A$ and $B$ label the two inequivalent sublattices spanning the graphene honeycomb lattice.
The envelope function sublattice components, $F^{\tau A}(\boldsymbol{r})$ and $F^{\tau B}(\boldsymbol{r})$, are just plane waves, provided the projection of the wave vector onto the direction perpendicular to the NT axis, $\varphi$, is properly quantized (Fig.~\ref{FIG:graphsheet}). 
This procedure, which corresponds to extract from the two-dimensional bands of graphene many one-dimensional sub-bands $\alpha$, leads to a graphene-like Dirac equation for the two-component envelope vector, $\boldsymbol{F}^{\tau}_{\alpha k  }$, 
\begin{eqnarray}
\label{eq.Dirac}
 \left( \begin{array}{cc}
   0 & \gamma k_{\tau}  - i \tau \gamma k \\
\gamma k_{\tau}  + i \tau \gamma k    & 0 \end{array}   \right)
\boldsymbol{F}^{\tau}_{\alpha k  } = E_{\alpha \tau  }(k)\, \boldsymbol{F}^{\tau}_{\alpha k  },
\end{eqnarray}
where $k_{\tau}$ is the quantized, transverse wave vector component, $k$ is the wave vector along the NT axis, and
$\tau=1$ for K, $\tau=-1$ for K$'$ valleys.
As we are interested in the long-range screening properties of narrow-gap NTs, out of all sub-bands $\alpha$ we consider only the top valence ($\alpha=c$) and bottom conduction ($\alpha=v $) bands closest to Dirac apexes.  
As shown below, this choice is validated a posteriori by comparing the dielectric function obtained from first principles with that obtained within the EM approximation as well as within the model of Sec.~\ref{sec:ourTheo}.
The dispersion of bands $c$ and $v$ is Dirac-like:
\begin{eqnarray}
\label{eq.eigenen}
E_{\alpha \tau  }(k) =  s_{\alpha} \gamma \sqrt{k_{\tau}^2+k^2}.
\end{eqnarray} 
Here $\gamma$ is graphene band parameter, $s_{\alpha}=1$ and $s_{\alpha}=-1$  for $c$ and $v$ bands, respectively.

\begin{figure}[b]
    \centering
    \includegraphics[clip,width=8.2cm]{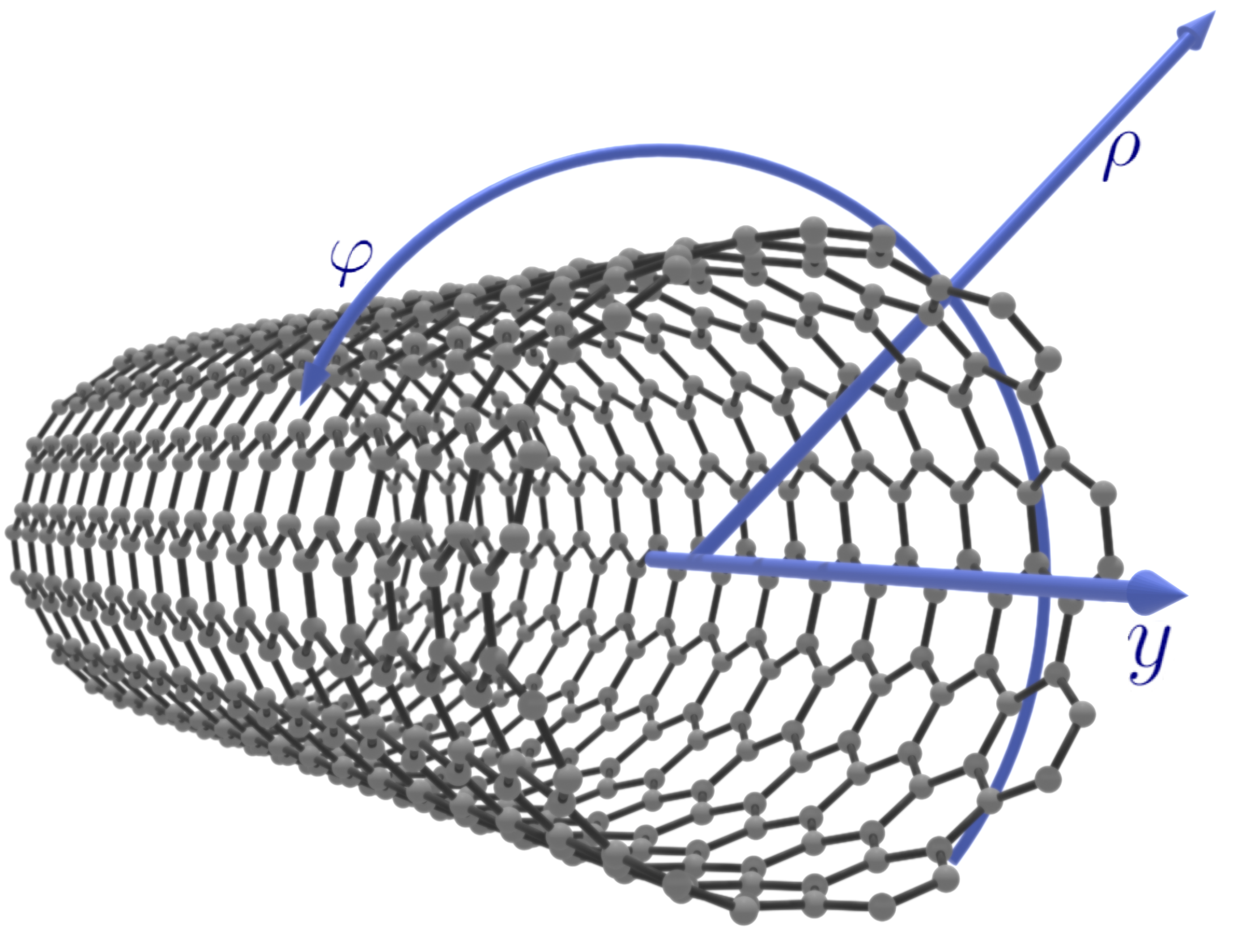}
    \caption{Pictorial representation of a nanotube with an armchair chirality.  Here $y$, $\varphi$, and $\rho$ are the directions of the nanotube axis, the direction on the circumference along which the folding has been performed, and the radial direction, respectively. The tube surface has radial coordinate $\rho=R$. }
    \label{FIG:armchair-figure}
\end{figure}

The solution of Dirac equation (\ref{eq.Dirac}) provides the phase relation between the two plane wave components of the envelope:
\begin{eqnarray}
\label{eq.envelope}
\! \boldsymbol{F}^{\tau}_{\alpha k } (\boldsymbol{r})= \left(  \begin{array}{c}  \! \! F^{A}_{\tau \alpha k  } \! \! \\ \! \! F^{B}_{\tau \alpha k  } \! \! \end{array}  \right) \! e^{i \boldsymbol{k} \cdot \boldsymbol{r}}\!=\! \frac{1}{\sqrt{2}} \! \left(  \begin{array}{c}  \! \frac{k_{\tau}-i \tau k}{\sqrt{k_{\tau}^2+k^2}} \! \\ \! s_{\alpha} \! \end{array}  \right) e^{i k y} e^{i k_{\tau} R \varphi}, 
\end{eqnarray}
where $R$ is the NT radius, $\varphi$ the azimuthal angle, and $y$ is the coordinate parallel to the NT axis, as shown in Fig.~\ref{FIG:armchair-figure}.

 The gap $2\gamma k_{\tau}$ we consider here is narrow,\cite{kane1997size,Charlier2007} usually ranging between 0 and 100 meV and hence smaller than the typical value characteristic of semiconducting NTs, of the order of 1 eV. This narrow gap originates from the curvature of the nanotube \cite{kane1997size} and may be tuned by an axial magnetic field through the Aharonov-Bohm effect,\cite{Ajiki1993} the two contributions to the gap adding in one valley and cancelling out in the other one.
 At zero field, the quantized wave vector $k_{\tau}$ is estimated as\cite{kane1997size,Charlier2007}
\begin{eqnarray}
\label{eq:smallgap}
k_{\tau} = \tau \frac{0.625 \ \text{eV}}{\gamma R^2} \cos(3 \theta),
\end{eqnarray}
with $\gamma=0.658$ eV nm. Here $\theta$ is the chiral angle identifying the direction along which the graphene is rolled, the zigzag and armchair orientations corresponding to $\theta=0$ and $\theta=\pi/6$, respectively (see Fig.~\ref{FIG:graphsheet}).  

\subsection{Dielectric function}
\label{effmod}

The EM dielectric function is built starting from the form that Coulomb interaction takes on a cylindrical surface\cite{Ando1997}:
\begin{eqnarray}
V_{\text{cyl}}(\boldsymbol{r},\boldsymbol{r'}) = \frac{e^2}{\sqrt{4 R^2 \sin^2\left(\frac{\varphi-\varphi'}{2 }\right) + (y-y')^2}}.
\end{eqnarray}
This potential may be expanded over azimuthal quantum number, $m$, and axial wave vector, $q$, components as 
\begin{multline}
V_{\text{cyl}}(\boldsymbol{r},\boldsymbol{r'}) = \frac{2 e^2}{A} \sum_{q}  \sum_{m=-\infty}^{\infty} \! \! \! I_{|m|}(q R)\, K_{|m|}(q R) \\ \times\quad e^{ i m (\varphi-\varphi')} \ e^{i q (y-y')}.\label{eq:Vcyl}
\end{multline}
Here $A$ is the nanotube length, while $I_m(x)$ and $K_m(x)$ are the modified Bessel functions of first and second kind, respectively. The RPA dielectric function, whose generic expression is given by 
Eq.~(\ref{RPA_1}), may then be recasted in terms of angular momenta $m$ rather than $\boldsymbol{G}$ vectors:
\begin{eqnarray}
\label{Eq.RPAem}
\epsilon_{\text{EM}}(m,q) & = & 1 - \frac{2 e^2}{A} I_{|m|}(q R)\, K_{|m|}(q R)\nonumber\\ && \times \quad \Pi_{\text{EM}}(m,q).
\end{eqnarray}
The polarisation $\Pi_{\text{EM}}$, which provides the independent-particle response, is written in terms of the wave functions of Eq.(\ref{eq.Bloch}):
\begin{align}
\notag \Pi_{\text{EM}}(m,q) & =  2 \sum_{\alpha, \alpha'} \sum_{\tau,\tau'} \sum_{k \in BZ} \! \! \langle \alpha \tau   k| e^{-i  q y } e^{-i  m \varphi } | \alpha'  \tau' k + q  \rangle  \\& \times\quad \notag \langle \alpha'  \tau'  k + q|  e^{i  q y' } e^{i  m \varphi' } | \alpha \tau   k \rangle \\& \quad\times\quad\frac{f(E_{\alpha' \tau'}(k + q)) - f(E_{\alpha \tau}(k))}{E_{\alpha' \tau'} (k+q) - E_{\alpha \tau  } (k)},
\end{align}
where the ket $| \alpha \tau   k \rangle$ is the NT orbital state $\Psi_{\alpha \tau   k}$ in Dirac notation \cite{Ando1997}.

A few simplifications are now in order. At zero temperature the difference between  the Fermi-Dirac factors, $f(E)$, selects virtual electron-hole excitations from filled valence to empty conduction band states only, hence the only relevant overlap integrals are $(\alpha=c, \alpha'=v )$ and $( \alpha=v, \alpha'=c)$. We ignore intervalley scatterings terms ($\tau \neq \tau'$), as they require large momentum transfer and are therefore negligible within the EM approximation.
Whereas in principle the sum over $k$ extends through the whole Brillouin zone, we truncate it through a cut-off  $k_o$, as done in Ref.~\cite{Ando1997}. We choose the cut-off to include only those $k$-points providing the bands with a Dirac-like shape, consistently with our previous work \cite{varsano2017carbon}. Eventually, by converting the sum over $k$ into an integral, one obtains
\begin{multline}
\label{RPA_2}
\epsilon_{\text{EM}}(m,q) =   1 - \frac{4 e^2}{\pi} I_{|m|}(q R)\, K_{|m|}(q R) \sum_{\alpha, \alpha'} \sum_{\tau} \int_{-k_o}^{k_o} \! \! \! d k \\ \times \quad  \langle \alpha \tau  k|  e^{-i q y} e^{-i m \varphi} | \alpha'  \tau k + q  \rangle   \langle \alpha' \tau k + q|  e^{i q y'} e^{i m \varphi'} | \alpha \tau   k \rangle \\ \times \quad \frac{f(E_{\alpha' k + q}) - f(E_{\alpha k})}{E_{\alpha' \tau'} (k+q) - E_{\alpha \tau  } (k)}.
\end{multline}


Importantly, the overlap integrals have the same form as those of graphene, the curved topology of the nanotube entering only through the quantized wave vector $k_{\tau}$:
 \begin{multline}
\label{eq.hom-overlap}
\langle \alpha k|  e^{-i  q y } e^{-i  m \varphi } | \alpha' k + q  \rangle =   
\\ \frac{1}{2} \left( \frac{k (k+q) +k_{\tau}^2-i q k_{\tau}}{\sqrt{k^2+k_{\tau}^2}\sqrt{(k+q)^2+k_{\tau}^2}} + (2 \delta_{\alpha,\alpha'} -1)  \right) \delta_{m,0} .
\end{multline}
Since only the $m=0$ angular momentum component is relevant within the 
two-band approximation, the dielectric function reduces to  
\begin{multline}
\epsilon_{\text{EM}}(q) =  1 + \frac{2e^2}{\pi \gamma} I_{0}(q R) K_{0}(q R)  \sum_\tau \Bigg[ \\ \frac{ \sqrt{(k_o + q)^2+ k_{\tau}^2} - \sqrt{(k_o - q)^2+ k_{\tau}^2} }{q} 
 +   \frac{ 2 k_{\tau}^2}{q\sqrt{q^2 + 4 k_{\tau}^2}} \\ \times\quad\log \Bigg(\frac{\sqrt{q^2 + 4 k_{\tau}^2} \sqrt{k_o^2+k_{\tau}^2} + 2 k_{\tau}^2 - k_o q} {\sqrt{q^2 + 4 k_{\tau}^2}\sqrt{(k_o + q)^2+ k_{\tau}^2} + 2k_{\tau}^2 + q(k_o + q)} \\ \times\quad \frac {\sqrt{q^2 + 4 k_{\tau}^2}\sqrt{(k_o - q)^2+ k_{\tau}^2} + 2k_{\tau}^2 + q(q-k_o)}{\sqrt{q^2 + 4 k_{\tau}^2} \sqrt{k_o^2+ k_{\tau}^2} + 2k_{\tau}^2 + k_o q} \Bigg) \Bigg],
 \label{eq:epsilonEM}
\end{multline}
which, in the limit of large cut-off $k_o$, simplifies to:
\begin{align}
\epsilon_{\text{EM}}(q) = & 1 + \frac{4 e^2}{\pi \gamma} I_{0}(q R) K_{0}(q R)  \sum_\tau \Bigg[ 1 \nonumber  \quad + \\  &  \frac{ 2 k_{\tau}^2}{q\sqrt{q^2 + 4 k_{\tau}^2}}   \log \Bigg(\frac{\sqrt{q^2 + 4 k_{\tau}^2} - q}{\sqrt{q^2 + 4 k_{\tau}^2} + q}\Bigg) \Bigg].
\end{align}

\section{Two-band model of screening}\label{sec:ourTheo}
 
In this section we improve the EM dielectric function by fully taking into account the three-dimensional topology of Bloch states $\psi_{\tau A/B} (\boldsymbol{r})$ that occur in the expression \eqref{eq.Bloch} for NT wave functions, while keeping the envelopes $F$ unchanged. The three-dimensional modelization of the Bloch states is illustrated in subsection \ref{ssec:3d}. We introduce (subsection \ref{ssec:supercell}) a large cylindrical supercell that contains the NT and then expand the states $\psi$ over the vectors $\boldsymbol{G}$ of the supercell three-dimensional reciprocal lattice. Here we avoid spurious interactions among supercell replicas by using the exact Coulomb cutoff technique of Refs.~\onlinecite{rozzi2006exact,ismail2006truncation}. The expressions for the dielectric function and dressed Coulomb interaction we obtain in subsection \ref{ssec:dielectric} exhibit an explicit dependence on reciprocal lattice vectors perpendicular to the NT axis, which accounts for the effect of tube curvature on wave functions.        

\subsection{Three-dimensional Bloch states}\label{ssec:3d}

The $\psi_{\tau A/B} (\boldsymbol{r})$ tight-binding Bloch states of Eq.~(\ref{eq.Bloch}) are localised on the atomic sites of the curved NT surface, whereas the EM model treats the lattice as two-dimensional. The position of these atoms depends in turn on the NT chirality, which may lead to a complex structure. For the sake of simplicity, we consider the exact atom location in two exemplar cases only, i.e., armchair and zigzag NTs, which are detailed, respectively, in Appendixes \ref{zigzag-ex-ov} and \ref{armchair-ex-ov}. Importantly, the forms of dielectric function and screened Coulomb potential that we obtain turn out to be identical to those derived from a simpler structural three-dimensional model \cite{Deslippe2009} that applies to all NT chiralities. Therefore, 
in this section we present only the model, which is validated in subsection \ref{sec:jellium} through comparison with the results for the true lattice.

\begin{figure}[t]
    \centering
    \includegraphics[width=8.6cm,trim={3cm 1cm 3cm 2cm},clip]{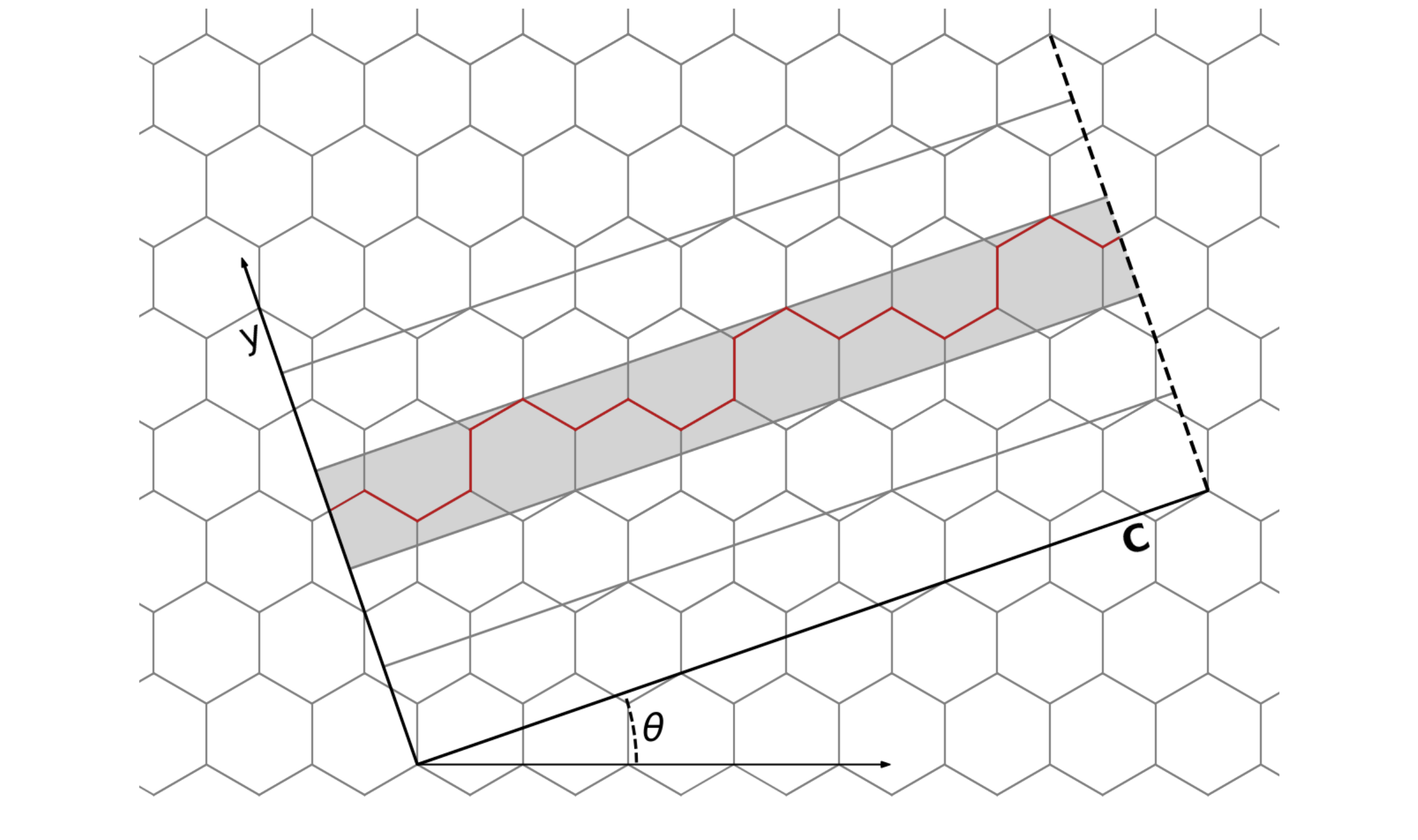}
    \caption{Pictorial illustration of the method used to build the homogeneous charge rings. $\boldsymbol{C}$ and $y$ are, respectively, the chiral vector and the nanotube axis. The dashed line on the right hand side signals where the cut of the graphene sheet is performed. We divide the nanotube surface into equally spaced stripes. Each stripe contains a unique closed chain of atoms (indicated in red in the highlighted stripe). The chains of atoms are replaced with two homogeneous charge rings, one for each sublattice. The rings are placed at the centre of the stripe.}
    \label{FIG:chain-figure}
\end{figure}

The model treats the NT structure as a series of $N$ rings over which the charge is spread homogeneously. As illustrated in Fig.~\ref{FIG:chain-figure}, the rings are perpendicular to the NT axis and their radius is equal to the tube radius, $R$. There are two species of rings, one for each sublattice. The ring positions along the $y$ axis,  
$\boldsymbol{R}^A_{l}=R^A_{l}\hat{y}$ or
$\boldsymbol{R}^B_{l}=R^B_{l}\hat{y}$, are given by:
\begin{align}
\Bigg\{ \begin{array}{c}
R^A_{l}=  \lambda l
+y^A_0 \\
R^B_{l}=  \lambda l 
+y^B_0
\end{array} \ \ \ \textrm{with} \ \ \ l= 1,...,N.
\end{align}
Here, $\lambda= a \cos \left(\pi/6-\theta \right)$ is the supercell length where $a=$ 0.246 $ \mbox{nm} $ is graphene lattice constant.
The rings are localised and equally spaced along the NT axis,
hence their $y$ coordinate may be thought of as an average over the positions of all atoms within a stripe of width $\lambda$ (shadowed area in Fig.~\ref{FIG:chain-figure}). Therefore, as the simplest possible approximation, we take the ring location at the origin to be the same for the two sublattices, i.e.,  
$y^A_0=y^B_0=0$.
The corresponding Bloch states are:
\begin{align}
\psi_{\tau A} (\boldsymbol{r})= \frac{1}{\sqrt{2 N}} \frac{e^{i \phi_{\tau A}}}{\sqrt{2 \pi R}}\sum_{l=1}^{N} \left[ e^{i \boldsymbol{K_\tau} \cdot \boldsymbol{R}^A_{l}} g (\boldsymbol{r}-\boldsymbol{R}^A_{l}) \right], \nonumber\\
\psi_{\tau B} (\boldsymbol{r})= \frac{1}{\sqrt{2 N}} \frac{e^{i \phi_{\tau B}}}{\sqrt{2 \pi R}} \sum_{l=1}^{N} \left[ e^{i \boldsymbol{K_\tau} \cdot \boldsymbol{R}^B_{l}} g (\boldsymbol{r}-\boldsymbol{R}^B_{l}) \right],
\label{eq:Blochjellium}
\end{align}
where $ \boldsymbol{K_\tau}$ is either K or K$'$, and  $\phi_{K A}=0$, $\phi_{K' A}=\theta$, $\phi_{K B}=-\frac{\pi}{3}+\theta$, $\phi_{K' B}=0$ (see Ref.~\onlinecite{secchi2010wigner}). The $g$ are functions localized on the tube surface, modeled as a homogeneous cylinder, whose square moduli behave as Dirac functions and which are defined as follows:
\begin{equation}
g^*\!(\boldsymbol{r}-\boldsymbol{R}^{\eta}_{l})\,g(\boldsymbol{r}-\boldsymbol{R}^{\eta'}_{l'}) =  \delta_{\eta,\eta'}  \delta_{l,l'}\,    \delta(\rho-R)\, \delta(y -
{R}^{\eta}_{l}),
\end{equation}
with $\rho$ being the radial coordinate. As the states of Eq.~\eqref{eq:Blochjellium} are achiral, NT orbitals $\Psi_{\alpha \tau   k}(\boldsymbol{r})$ depend on chirality solely through the curvature wave vector $k_{\tau}$ that enters the envelopes $\boldsymbol{F}^{\tau}_{\alpha k  }$.

\begin{figure}[b]
    \centering
    \includegraphics[clip,width=7.0cm]{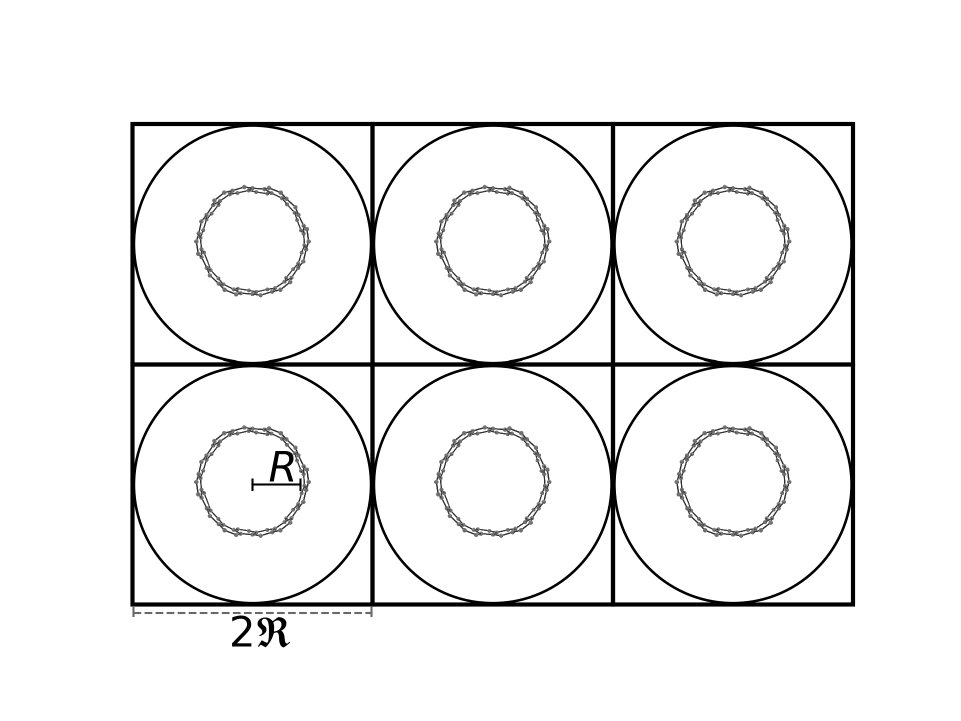}
    \caption{Sketch of the supercell structure in the non-periodic directions. Cylindrical supercells of radius $\mathfrak{R}$ are replicated and arranged on a square lattice. Each supercell contains a tube section, modeled as a ring of homogeneous charge, of radius $R$. }
    \label{FIG:supercell}
\end{figure} 

\subsection{Supercell calculation}\label{ssec:supercell}

In this subsection we mimic the approach from first principles
by building replicas of the tube along the directions perpendicular to the axis, as illustrated in Fig.~\ref{FIG:supercell}. As the whole system is now periodic, we may compute the polarisation $\Pi^{\text{CNT}}_{\boldsymbol{G},\boldsymbol{G'}}(\boldsymbol{q})$ as in  Eq.~\eqref{eq.polabin} through the three-dimensional plane-wave expansion, the reciprocal lattice vectors $\boldsymbol{G}$ depending on the size of the supercell containing a tube replica.
Throughout we use the acronym CNT to discriminate relevant quantities obtained in this section from the corresponding first-principles and EM results. 

Here we use a cylindrical supercell to contain the single NT unit (Fig.~\ref{FIG:supercell}). As the tube model structure is a sequence of rings along the axis, we identify a single ring as the building unit of the tube and hence allocate each ring of given axial coordinate $y$ in a different supercell. Thus, the length of the supercell $\lambda$ along the axis is equal to the distance between two subsequent rings, and the total length of the nanotube $A$ just amounts to $A=N \lambda$, where $N$ is the number of repetitions of the supercell along the axis. We work with a discretized set of axial wave vectors $q \rightarrow q_{j} $, where $q_{j} = 2 \pi j/A$ with $j= -N/2,..., N/2$. In the directions perpendicular to the nanotube axis we arrange the cylindrical supercells in a square superlattice with side equal to twice the radius of the supercell, $\mathfrak{R}$. We take $\mathfrak{R}$ to be much larger than $R$ to avoid quantum mechanical interactions among replicas.
Since the quantities of interest are obtained by sums over reciprocal lattice vectors, it is convenient to derive both axial and trasverse components, respectively $\boldsymbol{G}_{\parallel}$ and $\boldsymbol{G}_{\perp}$, in Cartesian form, from the periodic boundary conditions for the square superlattice: 
\begin{eqnarray}
\label{eq.G}
\boldsymbol{G}_{\perp} = \frac{\pi}{\mathfrak{R} } \left( n_1 \hat{x} + n_3 \hat{z} \right), \ \ \ \boldsymbol{G}_{\parallel} = \frac{2 \pi}{\lambda} n_2 \hat{y},
\end{eqnarray}
where $n_i=0,\pm 1,\pm 2,\ldots$, and $i=1,2,3$. The set of vectors over which we sum is determined through both a radial and an axial cutoff of the vector modulus, respectively $\left|\boldsymbol{G}_{\perp}\right|\le G_{\perp\text{max}}$ and $\boldsymbol{G}_{\parallel}\le G_{\parallel\text{max}}$, the error with respect to the usage of cylindrical coordinates being small in the limit of a dense set.  

In order to describe an isolated tube and hence avoid spurious Coulomb interactions among replicas of the system, which are due to the long range of the potential, we follow Ref.~\onlinecite{rozzi2006exact} and employ 
a form of the interaction that is truncated along the transverse directions:
\begin{multline}
\label{Coulomb}
v(\mathbf{q} + \mathbf{G}) =  
v_{\text{full}}(\mathbf{q} + \mathbf{G})
\Big[1 \quad + \\
+ \quad \mathfrak{R} \, G_{\perp}\, J_1(\mathfrak{R} G_{\perp}) \,K_0(\mathfrak{R} |q     + G_{\parallel}|) \quad + \\ -\quad   \mathfrak{R} \,|q  + G_{\parallel}|\, J_0(\mathfrak{R} G_{\perp})\, K_1(\mathfrak{R} |q + G_{\parallel}|)\Big].
\end{multline}
Here $v_{\text{full}}$ is the standard, bare three-dimensional Coulomb potential,
\begin{eqnarray}
\label{Coulomb-notruncation}
v_{\text{full}}(\mathbf{q} + \mathbf{G}) = \frac{4 e^2}{A \mathfrak{R}^2 (\mathbf{q} + \mathbf{G})^2},
\end{eqnarray}
$J_0(x), J_1(x)$ are Bessel functions of first kind, $K_0(x), K_1(x)$ are modified Bessel functions of second kind, and $\mathbf{q}=q \hat{y}$. In the case of armchair tubes only, which are gapless,
we use $v_{\text{full}}$ instead of $v$ as the full dressed interaction is cut-off in reciprocal space, and hence harmless.

The truncated potential $v$ oscillates in reciprocal space and is less divergent than $v_{\text{full}}$ at long wavelength, as Bessel functions $J_0(x), J_1(x)$ occurring in Eq.~\eqref{Coulomb} vanish with the argument $x$. As $v$ decreases quadratically with the magnitude of reciprocal lattice vectors, it is sufficient to include a limited number of $\boldsymbol{G}$ to reconstruct the Coulomb potential, either in real space [Eq.~\eqref{eq.Wtot} for the dressed potential $W$] or projected onto NT orbitals. Since the smallest $\boldsymbol{G_{\parallel}}$'s have magnitudes much larger than the first $\boldsymbol{G_{\perp}}$'s, the most relevant Fourier components are those with $\boldsymbol{G_{\parallel}}=0$ and $\boldsymbol{G_{\perp}}$ finite and small. In order to achieve convergence, both the supercell radius $\mathfrak{R}$ and the cutoffs $G_{\perp\text{max}}$ and $G_{\parallel\text{max}}$ must be carefully chosen, differing for the bare and screened Coulomb potential. The reconstruction of the bare potential requires large supercells and many $\mathbf{G}$ vectors, whereas the screened potential converges faster. For the dressed potential $W$, we take $\mathfrak{R}= 7R$ and $-15 \leq n_1, n_3 \leq 15$, including only the smallest finite axial vector $\boldsymbol{G_{\parallel}}$. 

\subsection{Dielectric function and dressed Coulomb potential}\label{ssec:dielectric}

The derivation of the polarisation $\Pi^{\text{CNT}}_{\boldsymbol{G},\boldsymbol{G'}}(\boldsymbol{q})$  
requires the knowledge of the overlap integrals $\rho_{cv}$ between $c$ and $v$ states
that occur in Eq.~\eqref{eq.polabin}. We compute these integrals by expanding
the Bloch states Eq.~\eqref{eq:Blochjellium} over the basis of three-dimensional plane waves with wave vector $\boldsymbol{G}+\boldsymbol{q}$, as detailed in Appendix \ref{appendix}.  
%
Explicitly, one has:
\begin{multline}
\label{epscnt}
 \epsilon^{\text{CNT}}_{\boldsymbol{G},\boldsymbol{G'}}(\boldsymbol{q}) =   \delta_{\boldsymbol{G},\boldsymbol{G'}} - \frac{2 A}{\pi} v(\mathbf{q} + \mathbf{G})  \sum_{\alpha, \alpha'} \sum_{\tau} \int_{-k_o}^{k_o} \! \!  d k \\  \langle \alpha \tau   k|  e^{-i (\boldsymbol{G} + \boldsymbol{q}) \cdot \boldsymbol{r}} | \alpha'  \tau k + q  \rangle   \langle \alpha'  \tau  k + q|  e^{i (\boldsymbol{G'} + \boldsymbol{q}) \cdot \boldsymbol{r'}} | \alpha \tau   k \rangle \\ \times \quad \frac{f(E_{\alpha' \tau'} (k + q)) - f(E_{\alpha \tau} (k))}{E_{\alpha' \tau'} (k+q) - E_{\alpha \tau  } (k)},
\end{multline}
with the overlap integrals being given by
\begin{multline}
\label{eq.ovecnt}
\langle \alpha k|  e^{-i (\boldsymbol{G} + \boldsymbol{q}) \cdot \boldsymbol{r}} | \alpha' k + q \rangle \quad =  \\ \left[ (F^{A}_{\tau  \alpha k})^{*} F^{A}_{\tau  \alpha' k+q} + (F^{B}_{\tau  \alpha k})^{*} F^{B}_{\tau \alpha' k+q}) \right] J_0(R G_{\perp}) \quad = \\ \frac{1}{2} \left( \frac{k (k+q) +k_{\tau}^2-i q k_{\tau}}{\sqrt{k^2+k_{\tau}^2}\sqrt{(k+q)^2+k_{\tau}^2}} + (2 \delta_{\alpha,\alpha'} -1) \right) J_0(R G_{\perp}). 
\end{multline}
Note that the overlap integral is similar to its EM counterpart Eq.~\eqref{eq.hom-overlap} except for the presence of the Bessel function $J_0$ of argument $RG_{\perp}$. This factor, absent in the EM expression, provides $\epsilon^{\text{CNT}}$ with the explicit dependence on tube curvature. 
After integration over $k$ and in the limit  $k_o \rightarrow \infty$ , the dielectric function reads:
\begin{align}
\notag 
\epsilon^{\text{CNT}}_{\boldsymbol{G},\boldsymbol{G'}} & (\boldsymbol{q})   =  \delta_{\boldsymbol{G},\boldsymbol{G'}} + \frac{2 A}{\pi \gamma} \  v(\boldsymbol{q} + \boldsymbol{G})\, J_0(R G_{\perp}) \, J_0(R G'_{\perp}) \\ &  \times \quad \sum_\tau \Bigg[ 1
 +  \frac{ 2 k_{\tau}^2}{q\sqrt{q^2 + 4 k_{\tau}^2}} \log\! \Bigg(\frac{\sqrt{q^2 + 4 k_{\tau}^2} - q}{\sqrt{q^2 + 4 k_{\tau}^2} + q}\Bigg) \Bigg] .
 \label{eq.epscnt}
\end{align}


In this work we focus on the matrix elements of the screened Coulomb interaction that bind electrons and holes together, mainly at small momentum transfer, $q$. Due to symmetry, electron-hole and electron-electron interaction have the same magnitude. The interaction matrix element, $W^{\tau}(k,k+q)$, is obtained by projecting the screened potential \eqref{eq.Wtot} over the electron-hole pair states $(c, \tau, k  )(v, \tau, k +q )$ and $(c,\tau, k + q)(v,\tau, k)$ within the same valley $\tau$:
\begin{align}
\label{eq:Wtcnt}
\notag W^{\tau}(k,k+q)= \sum_{\boldsymbol{G}} \sum_{\boldsymbol{G'}} \langle c \tau   k|  e^{-i (\boldsymbol{G'} + \boldsymbol{q}) \cdot \boldsymbol{r'}} | c \tau k+q  \rangle \\ \langle v \tau k+q |    e^{i (\boldsymbol{G} + \boldsymbol{q}) \cdot \boldsymbol{r}} | v \tau k \rangle \ (\epsilon_{\boldsymbol{G},\boldsymbol{G'}}(\boldsymbol{q}))^{-1} v(\boldsymbol{q} + \boldsymbol{G'}).
\end{align}
Since the corresponding first-principles quantity is evaluated on the grid $(k_{j}, k_{j}+q_l)$, it is convenient to integrate $W^{\tau}(k,k+q)$ over the reciprocal-space mesh $2\pi/A$. After inserting expressions \eqref{eq.epscnt} and \eqref{eq.ovecnt} into \eqref{eq:Wtcnt}, one obtains:
\begin{multline} 
\label{Wtcnt}
W^{\tau}_{\text{CNT}}(k_{j},k_{j}+q_{l})=\frac{A}{4 \pi} \int^{q_l + \pi/A}_{q_l - \pi/A} \hspace{-0.9cm}  dq \hspace{0.35cm}  \sum_{\boldsymbol{G}} \sum_{\boldsymbol{G'}} \! J_0(R G'_{\perp})\, J_0(R G_{\perp})  \\  [\epsilon^{\text{CNT}}_{\boldsymbol{G},\boldsymbol{G'}}(\boldsymbol{q})]^{-1}   v(\boldsymbol{q} + \boldsymbol{G'}) \left( 1 + \frac{k_j(k_j+q)+k_{\tau}^2} {\sqrt{k_j^2+k_{\tau}^2}\sqrt{(k_j+q)^2+k_{\tau}^2}} \right).
\end{multline} 
Note that integration regularizes the logarithmic singularity of Coulomb potential of Eq.~\eqref{Coulomb} for $q \rightarrow 0$, as
$ v(\mathbf{q} + \mathbf{G}_{\perp}) \simeq -\log(\mathfrak{R} \left|q\right|)$ for all allowed $G_{\perp}$'s.
For reference, the corresponding EM matrix element is:
\begin{multline} 
\label{WtEM}
W_{\text{EM}}^{\tau}(k_{j},k_{j}+q_{l})=\frac{e^2}{2 \pi} \int^{q_l + \pi/A}_{q_l - \pi/A} \hspace{-0.9cm}  dq \hspace{0.35cm} \epsilon_{\text{EM}}^{-1} (q) \ I_{0}(q R)\, K_{0}(q R) \\ 
\times\quad \left( 1 + \frac{k_j(k_j+q)+k_{\tau}^2} {\sqrt{k_j^2+k_{\tau}^2}\sqrt{(k_j+q)^2+k_{\tau}^2}} \right).
\end{multline}

\section{Results}
\subsection{Bare electron-hole interaction}\label{sec:bare}

\begin{figure}[b]
\centering
    \includegraphics[width=8.6cm,trim={0.1cm 0.0cm 0.0cm 0.4cm},clip]{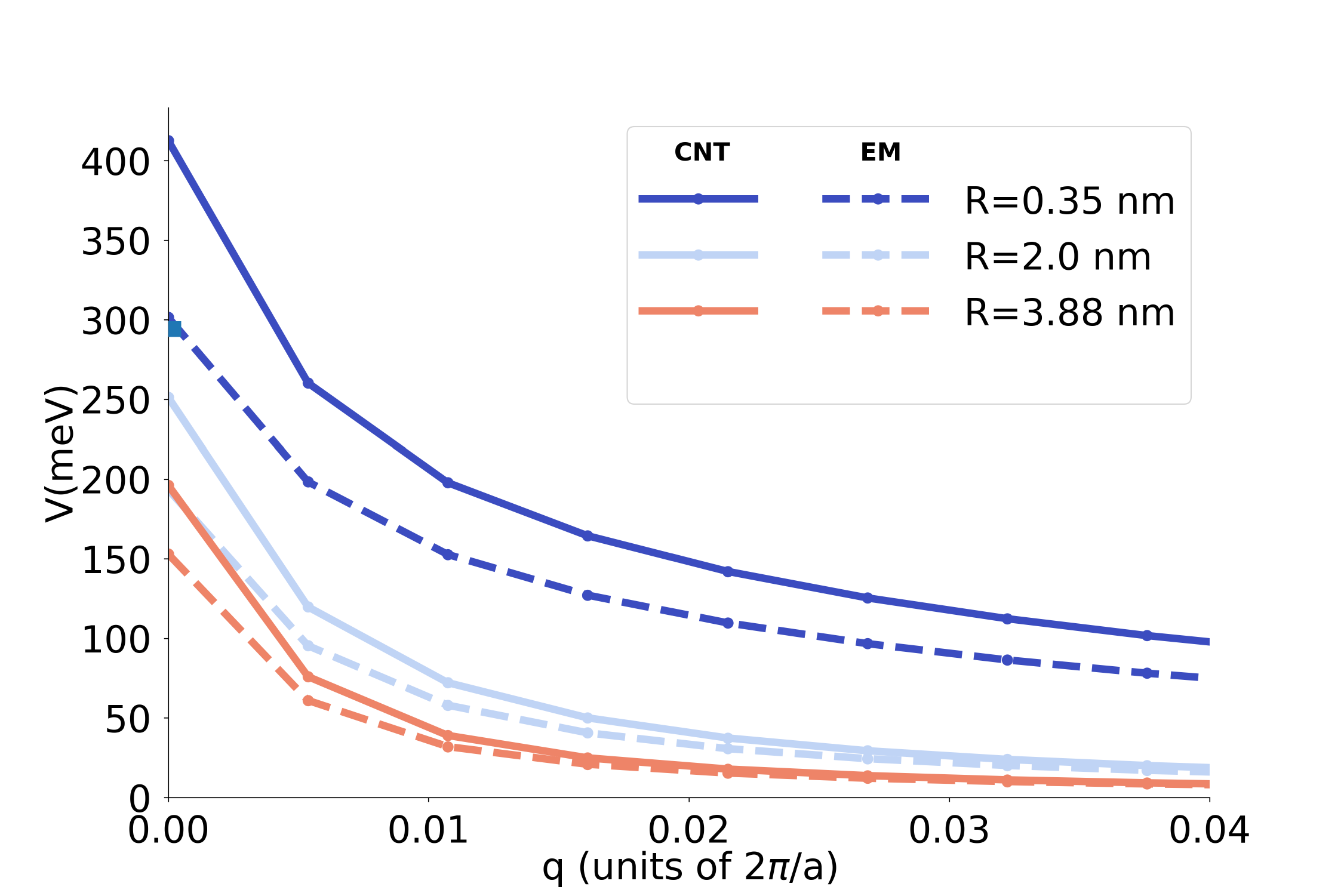}
\caption{Bare electron-hole interaction $V^{\tau}(q)$ vs $q$ computed from effective-mass theory (EM) and two-band model (CNT) for three  zigzag tubes of different radii $R$. The selected zigzag tubes are with increasing radii (9,0), (51,0) and (99,0). \label{Vnostro-Ando}}
\end{figure}

The key improvement of the two-band model of Sec.~\ref{sec:ourTheo} with respect to the effective-mass theory of Sec.~\ref{sec:EMwfs} shows up even in the absence of screening, when projecting the bare electron-hole interaction onto $c$ and $v$ Bloch states. This quantity, $V(q)$, is the matrix element $W^{\tau}(k_{j},k_{j}+q_{l})$ of equations \eqref{Wtcnt}
and \eqref{WtEM} evaluated for vanishing electronic polarisation, $\Pi=0$, that is $V^{\tau}(q_l)=\left[W^{\tau}(0,q_{l})\right]_{\Pi=0}$.   

We compare in Fig.~\ref{Vnostro-Ando} the two-band-model and EM matrix elements,
respectively $V^{\tau}_{\text{CNT}}(q_{l})$ and $V^{\tau}_{\text{EM}}(q_{l})$, for different NT radii $R$.
Recall that the numerical discretization of momentum space regularizes the logarithmic singularity expected for $q\rightarrow 0$.
The two-band-model bare electron-hole interaction systematically exceeds its EM counterpart, as only the former is sensitive to the curved tube topology.
The enhancement of the interaction originates
from the the form factors of the kind $J_0(R G_{\perp})$ that modify graphene overlap integrals. The mismatch between $V_{\text{CNT}}$ and $V_{\text{EM}}$ is stronger for smaller $R$ and softens as the tube curvature becomes negligible.  

Note that $V_{\text{CNT}}$ and $V_{\text{EM}}$ build on different expression of the full, non-projected Coulomb potential, depending respectively on $\boldsymbol{G}$ vectors and azimuthal quantum numbers $m$. However, the two potential forms, once evaluated on the same cylindrical surface of radius $R$, are identical at long wavelength in the macroscopic limit, $v \sim 2e^2/A\log (A/R)$, as we show explicitly in Appendix \ref{Coul-analysis}.   


\subsection{Dielectric function}\label{sec:die}

\begin{figure*}[t!]
\hspace*{-3mm}
\begin{tabular}{cccc}
\text{(a)} & & \hspace{0.5cm} \text{(b)} & \\
& \hspace{-0.5cm} \includegraphics[width=8.6cm,trim={0.5cm 0.2cm 0.0cm 0.0cm},clip]{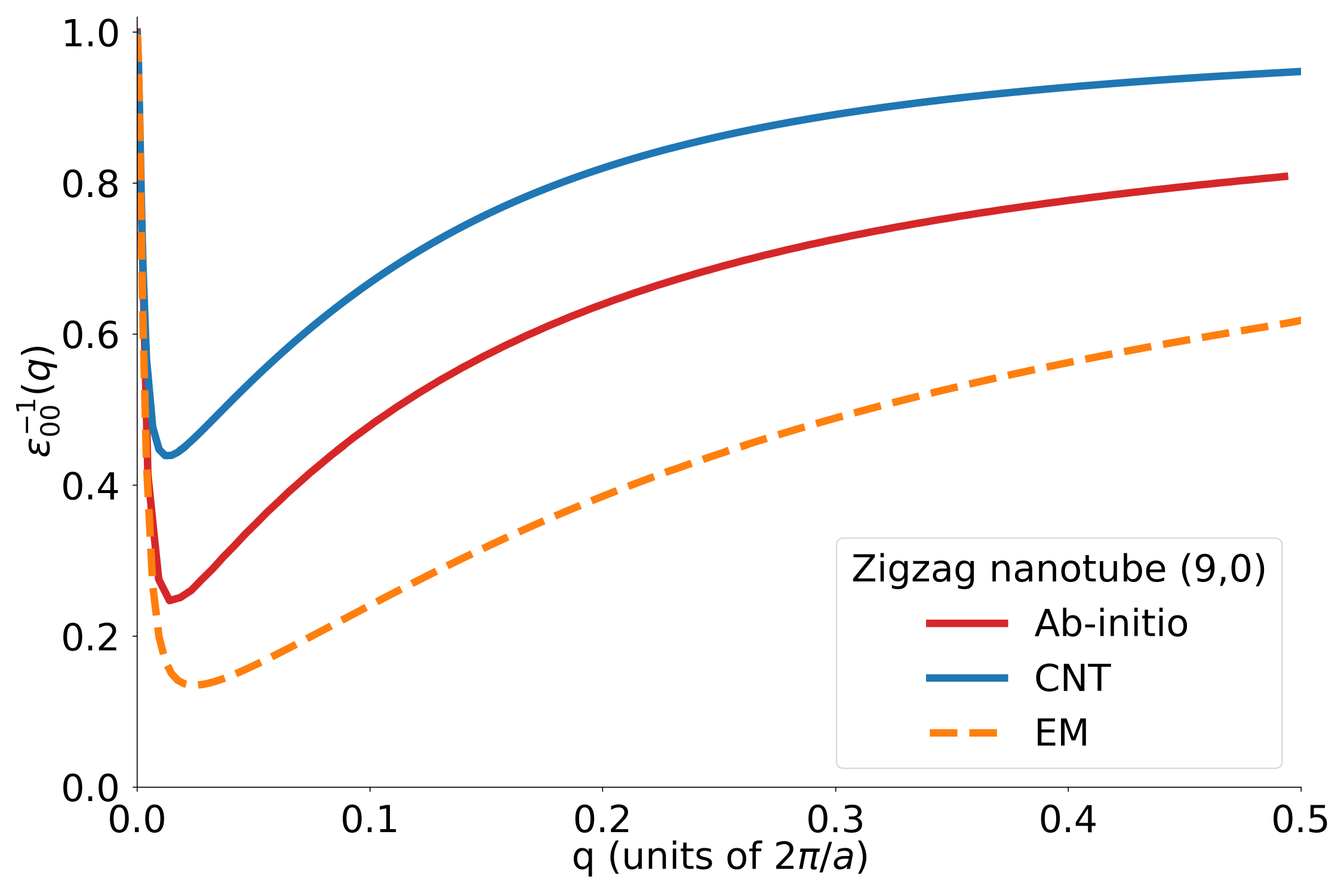}  &   & \hspace{-0.5cm} \includegraphics[width=8.6cm,trim={0.5cm 0.2cm 0.0cm 0.0cm},clip]{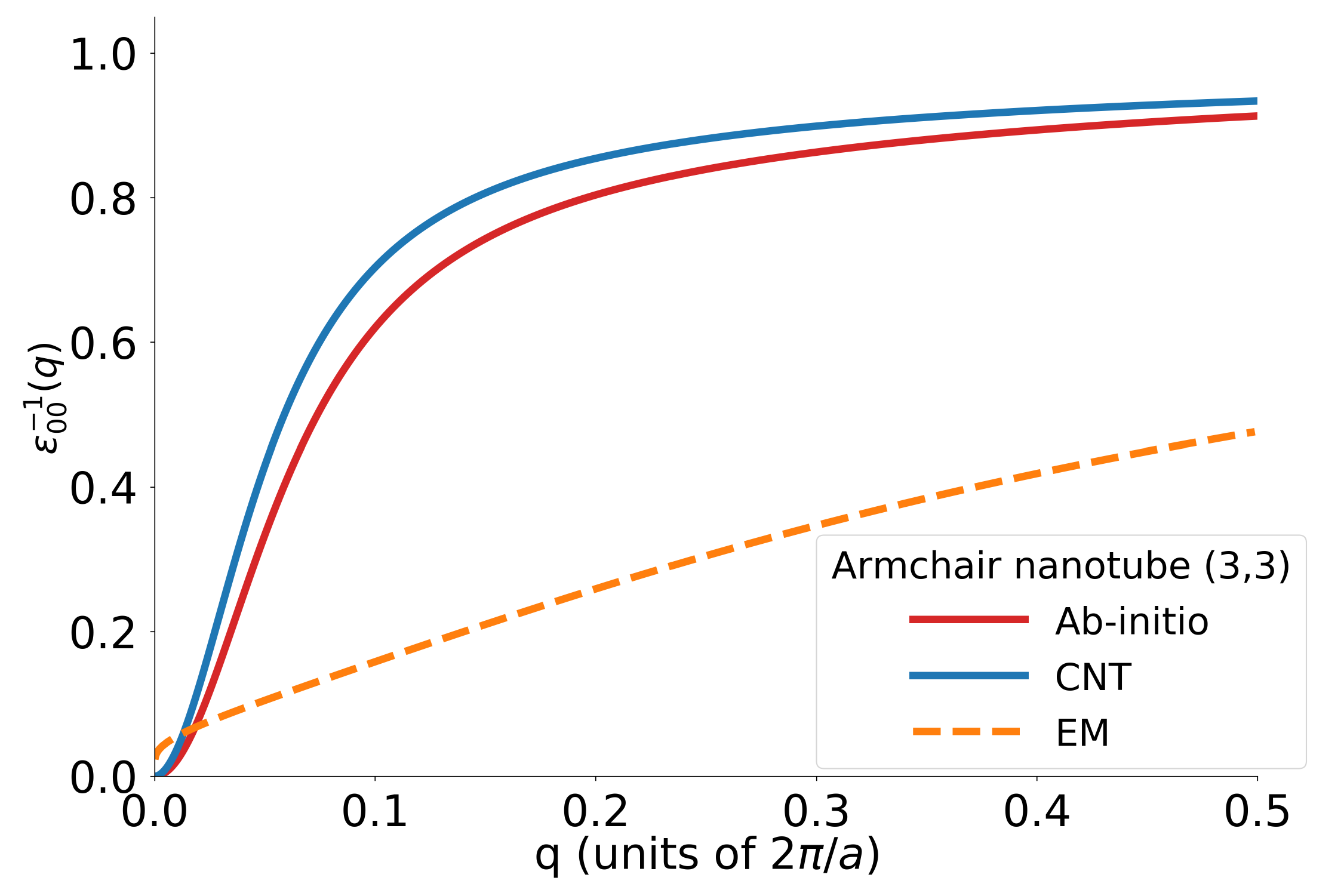} \
\end{tabular}
\caption{Macroscopic dielectric function $\epsilon^{-1}_{0,0}(q)$ vs $q$ computed from first principles (ab initio), effective-mass theory (EM), and two-band model (CNT). Panels (a) and (b) show data for zigzag (9,0) and armchair (3,3) nanotubes, respectively. 
\label{FIG.epss}}
\end{figure*}

\begin{figure*}[t!]
\hspace*{-3mm}
\begin{tabular}{cccc}
\text{(a)} & & \hspace{0.5cm} \text{(b)} & \\
& \hspace{-0.5cm}
\includegraphics[width=8.6cm,scale=0.2,trim={0.5cm 0.2cm 0.0cm 0.0cm},clip]{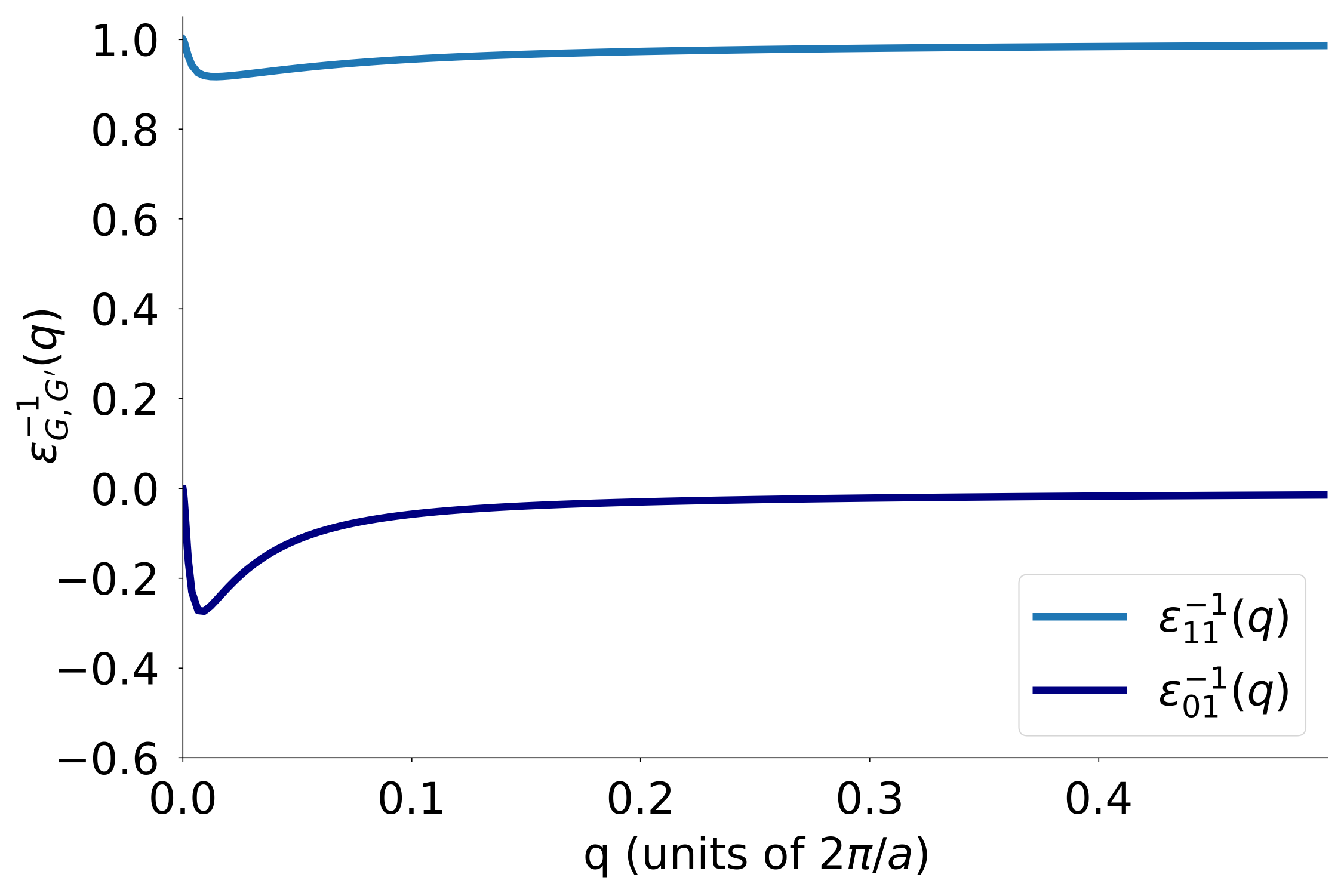} & & \hspace{-0.5cm} \includegraphics[width=8.6cm,scale=0.2,trim={0.5cm 0.2cm 0.0cm 0.0cm},clip]{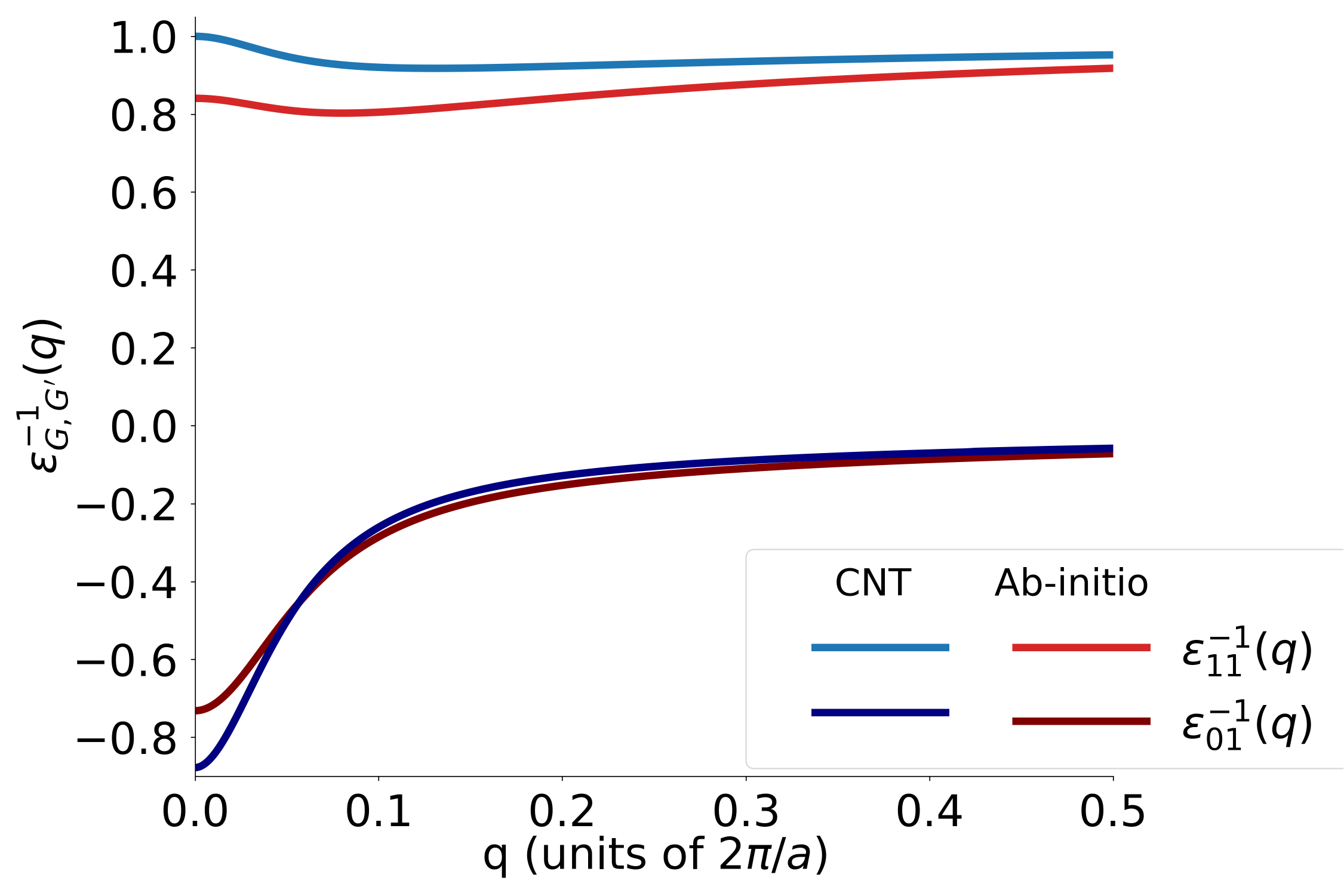}
\end{tabular}
\caption{Diagonal, $\epsilon^{-1}_{\boldsymbol{G}_{\perp},\boldsymbol{G}_{\perp}}(\boldsymbol{q})$, and wing term, 
$\epsilon^{-1}_{0,\boldsymbol{G}_{\perp}}(\boldsymbol{q})$,
of the inverse dielectric matrix
vs momentum $q$ for the smallest vector $\boldsymbol{G}_{\perp}$ with $n_1=1$, $n_2=n_3=0$.
Panel (a): Two-band-model results for the (9,0) zigzag nanotube. Panel (b): first-principles (ab initio) and two-band-model (CNT) results for the (3,3) armchair nanotube.   
\label{eps-micro}}
\end{figure*}

Large-gap semiconducting carbon nanotubes are known to poorly screen charge carriers at electron-electron separations that are either large or small with respect to the NT radius, as an effect of the low dimensionality.\cite{leonard2002dielectric,Deslippe2009} Our calculations from first principles for narrow-gap NTs show a qualitatively similar behaviour, provided one replaces the crossover length $R$ with $\left|k_{\tau}\right|^{-1}$.

Figure \ref{FIG.epss} (a) reports the dependence of the inverse ``macroscopic'' dielectric function $\epsilon^{-1}_{0,0}(\boldsymbol{q})$ on the wave vector $q$ for the (9,0) zigzag NT (red curve), whose calculated gap is 110 meV. For both small and large $q$ the inverse dielectric constant is close to one, the crossover occurring close to $q\approx \left|k_{\tau}\right|= 6\times 10^{-3}$ $2\pi/a$. The trend of $\epsilon^{-1}$ of is qualitatively similar to that of large-gap NTs, like the (8,0) tube shown in Fig.~1a of Deslippe {\it et al.}\cite{Deslippe2009}, except for the different crossover location. 
The rationale is that, for large-gap semiconducting NTs, the ``secondary'' contribution to the gap, due to curvature and proportional to $\left|k_{\tau}\right|$ as defined in Eq.~\eqref{eq:smallgap}, is negligible with respect to the ``primary gap'' proportional to $1/R$, whereas in narrow-gap NTs the primary gap is absent.\cite{Charlier2007} The gapless limit of armchair tubes is regained for $k_{\tau}\rightarrow 0$, which allows for metallic screening at long wavelength, i.e.,
$\epsilon^{-1}_{0,0}(q=0)=0$. This is shown for the (3,3) tube by the red curve of Fig.~\ref{FIG.epss} (b), which exactly reproduces Fig.~2 of Spataru {\it et al.}\cite{Spataru2004b}. This result, which builds on the full bare potential $v_{\text{full}}$, is cell-independent and hence may be used a benchmark for model approaches, whereas $\epsilon^{-1}_{0,0}(\boldsymbol{q})$ of panel (a) depends on the supercell size.  

The two-band model calculation of $\epsilon^{-1}_{0,0}$  (blue curves in Fig.~\ref{FIG.epss}, CNT) reproduces quantitatively the inverse dielectric constant of the armchair tube from first principles, the difference between ab-initio and CNT curves remaining small in the whole $q$ range. On the other hand, a direct comparison with the zigzag tube is not possible, due to the size mismatch between first-principles and model supercells, which affects the magnitude of the macroscopic bare truncated potential $v$ and hence $\epsilon^{-1}_{0,0}$. The systematic enhancement of the model result with respect to first-principles data is likely due to the neglect of higher-energy virtual electron-hole excitations, which are responsible for the screening effect. 

Contrary to the model prediction, the EM calculation of the inverse dielectric constant performs poorly for the armchair tube [dashed curve in Fig.~\ref{FIG.epss}(b)], even failing to reproduce the correct curvature of $\epsilon^{-1}_{0,0}(q)$ at $q\approx 0$ and grossly missing its magnitude.
Regardless of chirality, the EM theory overestimates substantially the electronic polarization with respect to the two-band model.

We have checked that the non-local terms of the inverse dielectric matrix
$\epsilon^{-1}_{\boldsymbol{G},\boldsymbol{G'}}(\boldsymbol{q})$ that
have finite transverse vectors $\boldsymbol{G}_{\perp}$ strongly affect the dressed electron-hole interaction $W$. The most relevant terms turn out to be the diagonal matrix elements of kind
$\epsilon^{-1}_{\boldsymbol{G}_{\perp},\boldsymbol{G}_{\perp}}(\boldsymbol{q})$ and the ``wing'' terms of type $\epsilon^{-1}_{\boldsymbol{G}_{\perp},0}(\boldsymbol{q})$ [or
$\epsilon^{-1}_{0,\boldsymbol{G}_{\perp}}(\boldsymbol{q})$].
For the sake of illustration, Fig.~\ref{eps-micro} shows the dependence of the first diagonal and wing matrix elements of $\epsilon^{-1}$  on momentum $q$ for selected tubes. The shown trend is generic for all vectors $\boldsymbol{G}_{\perp}$, the model and first-principles results being almost identical. The diagonal elements are close to unity and thus enhance the dressed interaction, whereas the wing terms are small and negative, thus increasing the screening effect. As clear from Eq.~\eqref{Wtcnt}, as the magnitude of $\boldsymbol{G}_{\perp}$ increases the weight of its contribution to $W$ decreases approximately as $\left|\boldsymbol{G}_{\perp}\right|^{-2}$. 

At long wavelength, gapped and gapless tubes behave differently.
As shown for the gapped zigzag (9,0) tube in panel (a), both diagonal and wing terms of the inverse dielectric constant  exhibit a minimum close to $q\approx \left|k_{\tau}\right|$, like the macroscopic term $\epsilon^{-1}_{0,0}$ of Fig.~\ref{FIG.epss}(a), corresponding to a maximum of the polarisation $\Pi$. 
For $q\rightarrow 0$ the polarisation vanishes quadratically, 
as apparent from the analytical behaviour of the model polarisation (only valid in the presence of the gap),
\begin{align}&
\Pi^{\text{CNT}}_{\boldsymbol{G},\boldsymbol{G'}}(\boldsymbol{q}\simeq 0) = - \frac{2 A}{3 \pi \gamma k_{\tau}^2} \  q^2 J_0(R G_{\perp}) J_0(R G'_{\perp}). 
\label{eq:curvature}
\end{align}
Therefore,
the diagonal (wing) term tends to unity (zero). For gapless tubes,
like the (3,3) armchair tube of Fig.~\ref{eps-micro}(b), the maximum of $\Pi$ moves to $q=0$ together with the mimimum of wing terms, whereas the diagonal terms weakly depend on $q$.

\subsection{Screened electron-hole interaction}\label{sec:dressed}

The key quantity we focus on is the screened, momentum-dependent electron-hole interaction, $W(q)$, projected on $c$ and $v$ bands. This matrix element provides the Bethe-Salpeter equation of motion for excitons with the non-trivial information about screening. Furthermore, the
dressed interaction obtained from the two-band and EM approaches, $W(q)=W^{\tau}(0,
q_l)$ as defined in equations \eqref{Wtcnt}
and \eqref{WtEM}, may be directly compared with the matrix element obtained from first principles, for given sampling of momentum space.
In this subsection we discuss gapped NTs and postpone the gapless case to subsection \ref{sec:armchair}, as the latter case requires special handling in view of its singular behaviour in the limit $q\rightarrow 0$, $k_{\tau}\rightarrow 0$.

\begin{figure}[t!]  
\begin{tabular}{cc}
\includegraphics[width=8.6cm,trim={0.5cm 0.5cm 0.0cm 0.0cm},clip]{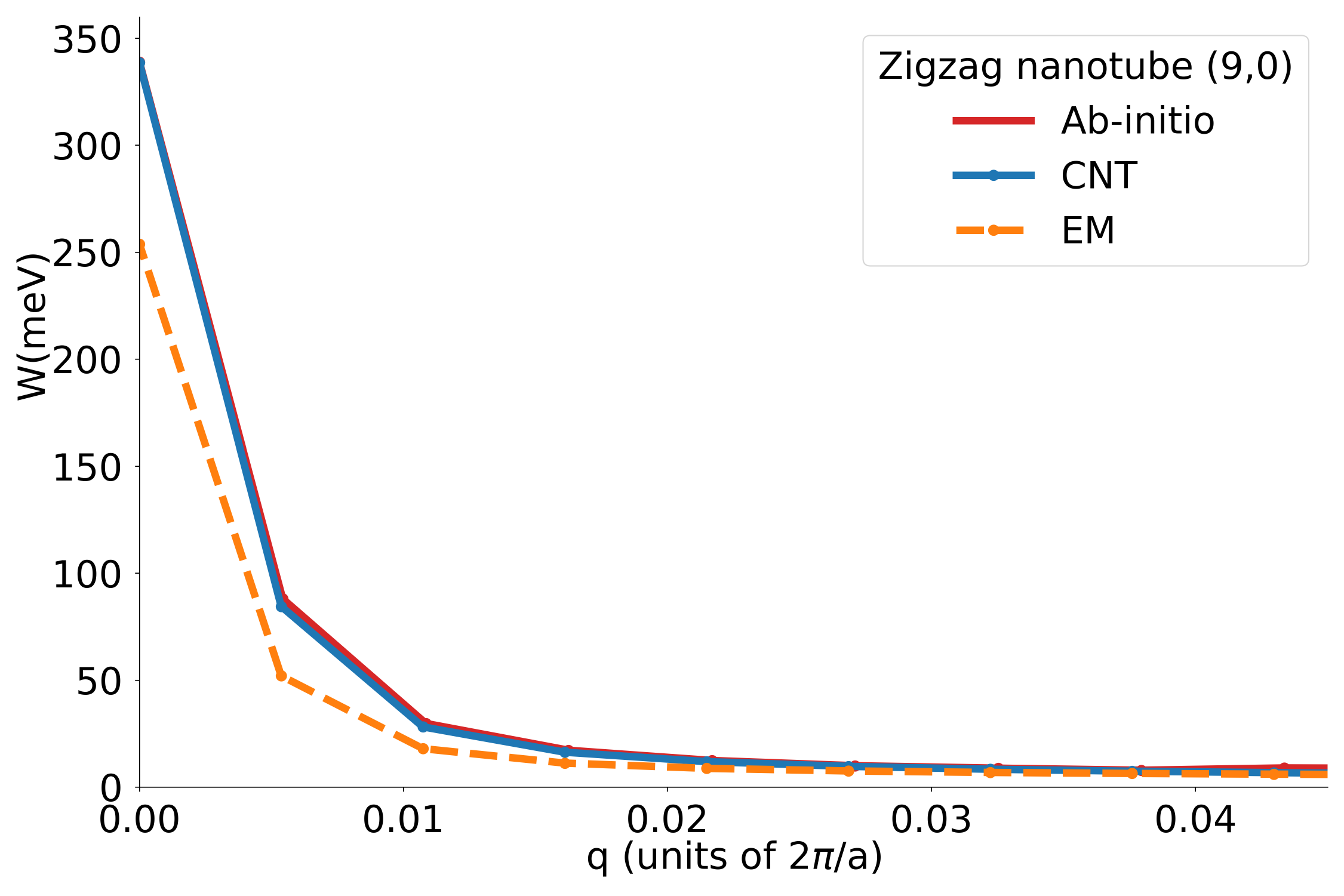} \\
\end{tabular}
\caption{Screened electron-hole interaction $W(q)$ vs $q$ for the zigzag (9,0) nanotube derived from first-principles (ab-initio),  effective mass (EM) and  and two-band model (CNT) approaches. } \label{potYambo}
\end{figure}

\begin{figure}[t]
\includegraphics[width=8.6cm,trim={0.0cm 0.0cm 0.0cm 0.0cm},clip]{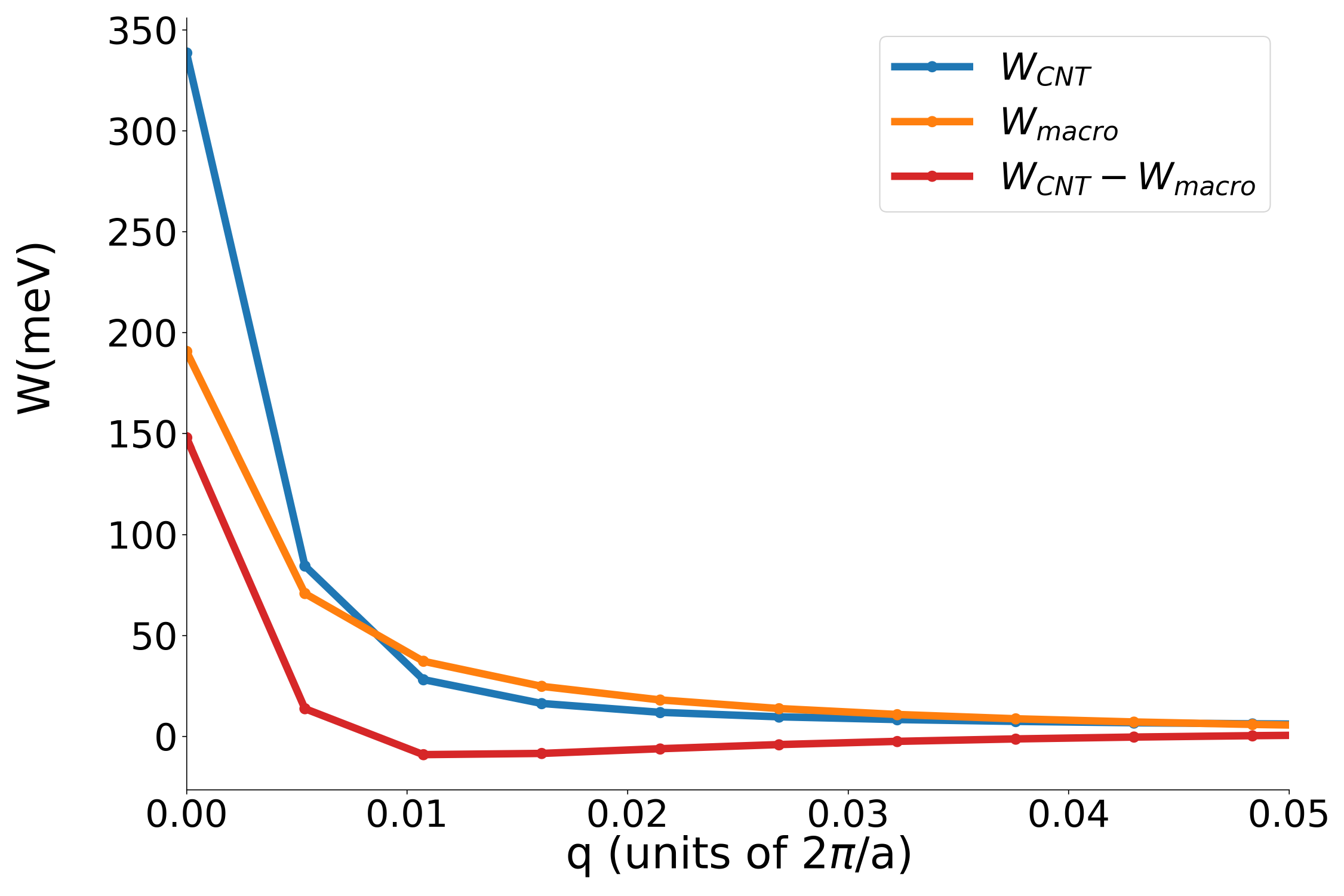}
\centering
\caption{Macroscopic and microscopic contribution to the model dressed electron-hole interaction, $W(q)$ vs $q$, for the zigzag (9,0) nanotube.
In the sum over terms depending on $\epsilon^{-1}_{\boldsymbol{G}_{\perp},\boldsymbol{G'}_{\perp}}(\boldsymbol{q})$, the macroscopic term ($W_{macro}$) corresponds to $(\boldsymbol{G}_{\perp},\boldsymbol{G'}_{\perp})=(0,0)$ and the microscopic term ($W_{CNT}-W_{macro}$) is the remainder.  \label{macromicro_90}}
\end{figure}

As illustrated by Fig.~\ref{potYambo}, the two-band-model calculation of $W_{\text{CNT}}$ (blue curve) agrees very well with first-principles data (red curve) for the zigzag (9,0) tube. On the contrary, EM theory (dashed curve) substantially overestimates screening at small momentum transfer, and hence invariably underestimates exciton binding energies. The key to the
perfect matching of first-principles and model approaches is the full inclusion of local-field effects, as illustrated by the model calculation of Fig.~\ref{macromicro_90}. Here we separate the ``macroscopic'' and ``microscopic'' contributions to $W(q)$ of equation \eqref{Wtcnt} in the sum over terms proportional to $\epsilon^{-1}_{\boldsymbol{G}_{\perp},\boldsymbol{G'}_{\perp}}(\boldsymbol{q})$, where the former is term $(\boldsymbol{G}_{\perp},\boldsymbol{G'}_{\perp})=(0,0)$ and the latter is the remainder of the sum. The macroscopic term provides $W$ with the gross contibution, but local-field terms are essential to regain the actual potential. Whereas diagonal terms $(\boldsymbol{G}_{\perp},\boldsymbol{G}_{\perp})$ increase the interaction strength and are most effective at $q\approx 0$, the
wing terms $(\boldsymbol{G}_{\perp},0)$ enhance screening and are most relevant  for $q > \left|k_{\tau}\right|$, where 
the microscopic contribution
(labeled $W_{CNT}-W_{macro}$) becomes negative. 

\begin{figure}[t!]
\hspace*{-3mm}
\centering
\includegraphics[width=8.6cm,height=0.43\textwidth,trim={0.1cm 0.0cm 0.0cm 0.4cm},clip]{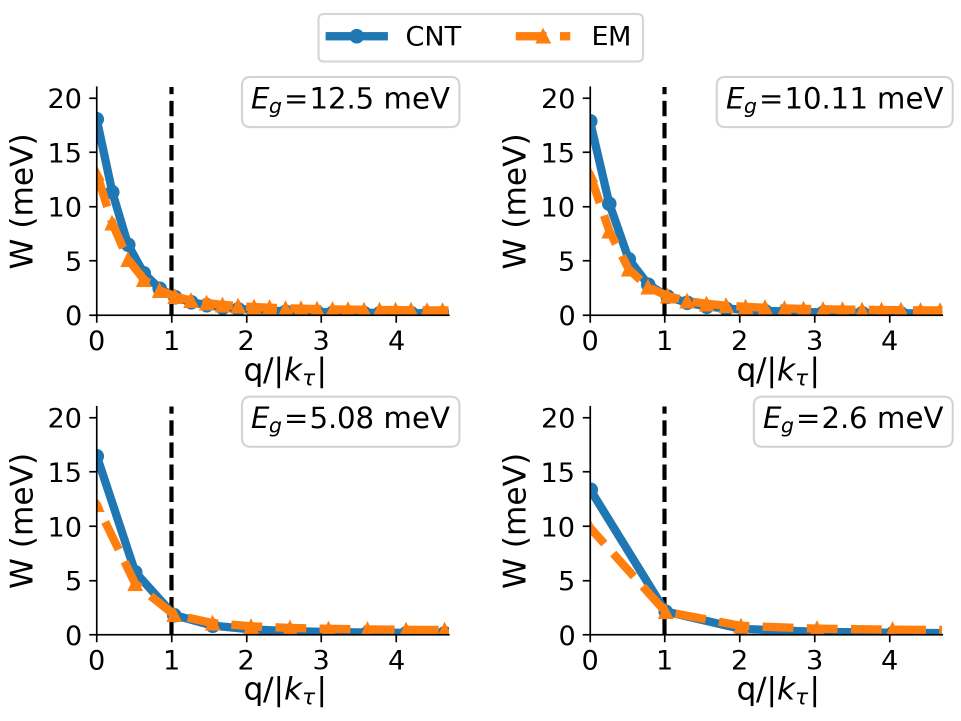}
\caption{Screened electron-hole interaction potential, $W(q)$, vs renormalized momentum, $q/\left|k_{\tau}\right|$, from model and effective-mass calculations, for different gap values, $E_g$. The nanotube radius is fixed, $R=1$ nm, and the vertical dashed line corresponds to $q=\left|k_{\tau}\right|$. \label{R=1_1}}
\end{figure}

\begin{figure}[t!]
\hspace*{-3mm}
\centering
\includegraphics[width=8.6cm,height=0.43\textwidth,trim={0.1cm 0.0cm 0.0cm 0.0cm},clip]{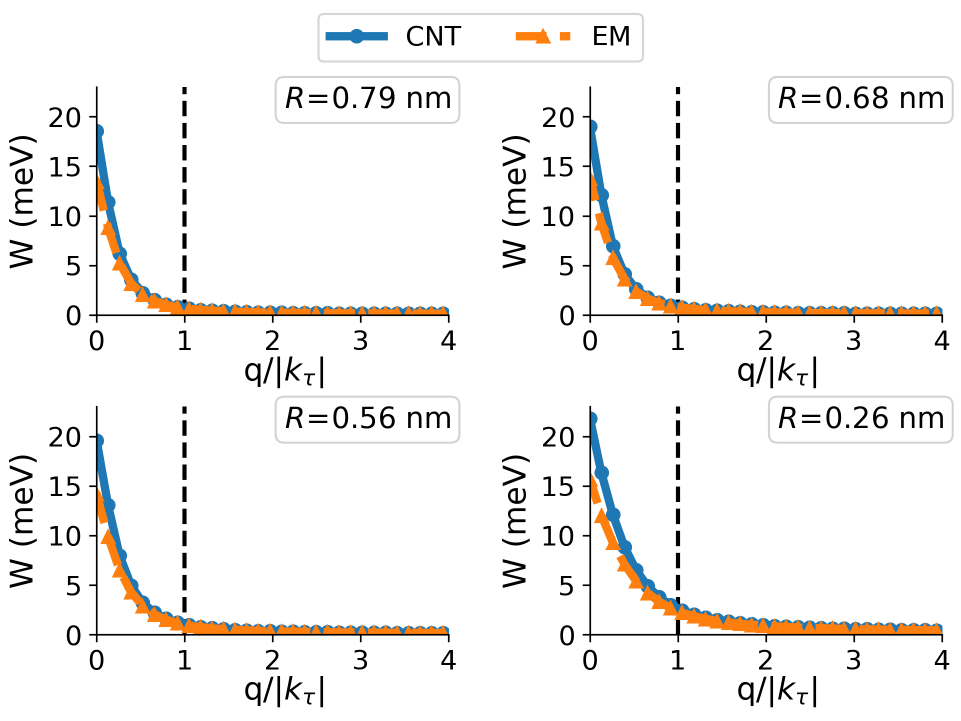}
\caption{
Screened electron-hole interaction potential, $W(q)$, vs renormalized momentum, $q/\left|k_{\tau}\right|$, from model and effective-mass calculations, for different nanotube radii, $R$. The nanotube gap is fixed, $E_g=$ 20 meV, and the vertical dashed line corresponds to $q=\left|k_{\tau}\right|$.
 \label{Eg=20_1}}
\end{figure}

The model calculation allows to derive the screened electron-hole interaction for tubes of any radius, $R$, and gap size, $E_g$, the latter being fixed by the combination of $R$ and $\theta$ given in equation \eqref{eq:smallgap}.    
This is illustrated for selected sets of $(R,E_g)$ values by Figs.~\ref{R=1_1} and \ref{Eg=20_1} (here we treat $R$, $E_g$, and $\theta$ as continuous parameters). 
We fix either the radius
($R=$ 1 nm in Fig.~\ref{R=1_1}) or the gap size ($E_g=$ 20 meV in Fig.~\ref{Eg=20_1}) and plot
the dressed interaction $W(q)$ as a function of the renormalized
momentum $q/\left|k_{\tau}\right|$. 
All plots of $W$ exhibit an almost identical dependence on
$q/\left|k_{\tau}\right|$, which demonstrates that the most relevant length scale is $\left|k_{\tau}\right|^{-1}$, whereas the absolute magnitude of $W$ at long wavelength, $W(q=0)$, depends in a non-trivial way on both $R$ and $\left|k_{\tau}\right|^{-1}$. 
In particular, $W(q=0)$ decreases weakly with $R$ for given energy gap (Fig.~\ref{Eg=20_1}), whereas the bare interaction $V(q=0)$ substantially depends on $R$ (Fig.~\ref{Vnostro-Ando}). Thus, screening tends to weaken the dependence of the dressed potential on $R$ and to enhance that on   $\left|k_{\tau}\right|^{-1}$.
A key result is that, 
for $q<\left|k_{\tau}\right|$, EM and two-band-model predictions systematically depart, the EM approximation substantially overestimating screening.

\subsection{Armchair tubes and excitonic instability}\label{sec:armchair}

As the gap vanishes, as in armchair NTs, screening acquires a metallic character, becoming effective even at long wavelength. As a consequence,
the electronic polarization $\Pi$ exhibits a non-analytic behaviour in the limit $q\rightarrow 0$, $k_{\tau}\rightarrow 0$. This is illustrated by the quadratic expansion 
 of $\Pi$ (Eq.~\ref{eq:curvature}) for small $q$ values, which tends to zero
or infinity depending on the order of the limits $\lim_{q\rightarrow 0}$
and $\lim_{k_{\tau}\rightarrow 0}$. 

\begin{figure}[h]
\includegraphics[width=8.6cm,trim={0.5cm 0.0cm 0.0cm 0.0cm},clip]{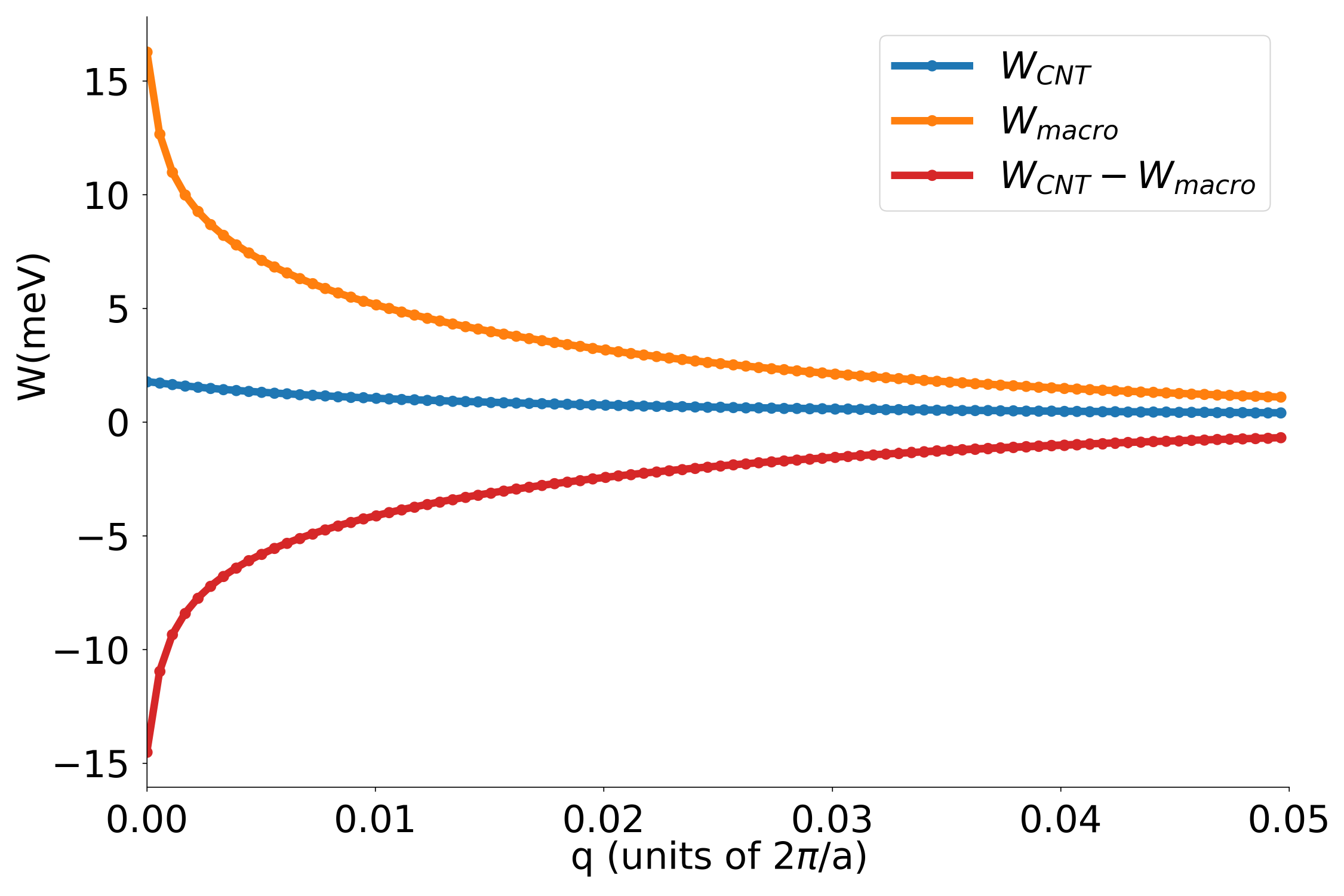}
\centering
\caption{
Macroscopic and microscopic contribution to the model dressed electron-hole interaction, $W(q)$ vs $q$, for the armchair (5,5) nanotube.
In the sum over terms depending on $\epsilon^{-1}_{\boldsymbol{G}_{\perp},\boldsymbol{G'}_{\perp}}(\boldsymbol{q})$, the macroscopic term ($W_{macro}$) corresponds to $(\boldsymbol{G}_{\perp},\boldsymbol{G'}_{\perp})=(0,0)$ and the microscopic term ($W_{CNT}-W_{macro}$) is the remainder.
\label{macromicro_55}}
\end{figure}

A counterpart to the singularity of $\Pi$ is the enhanced role of microscopic local fields in the building of the dressed electron-hole interaction, as shown in Fig.~\ref{macromicro_55}
for the armchair (5,5) tube. Here the microscopic contribution ($W_{CNT}-W_{macro}$) to $W$ 
is large and negative for $q\rightarrow 0$, due to dominance of wing terms $(\boldsymbol{G}_{\perp},0)$
in the sum over $(\boldsymbol{G}_{\perp},\boldsymbol{G'}_{\perp})$, sensitive to the magnitude of $\Pi$. This enhancement leads to a major cancellation of the macroscopic term $(\boldsymbol{G}_{\perp},\boldsymbol{G'}_{\perp})=(0,0)$, which has opposite sign and comparable magnitude ($W_{macro}$ in Fig.~\ref{macromicro_55}), and hence requires careful numerical handling. 

\begin{figure}[h]
\includegraphics[width=8.6cm,trim={0.5cm 0.0cm 0.0cm 0.0cm},clip]{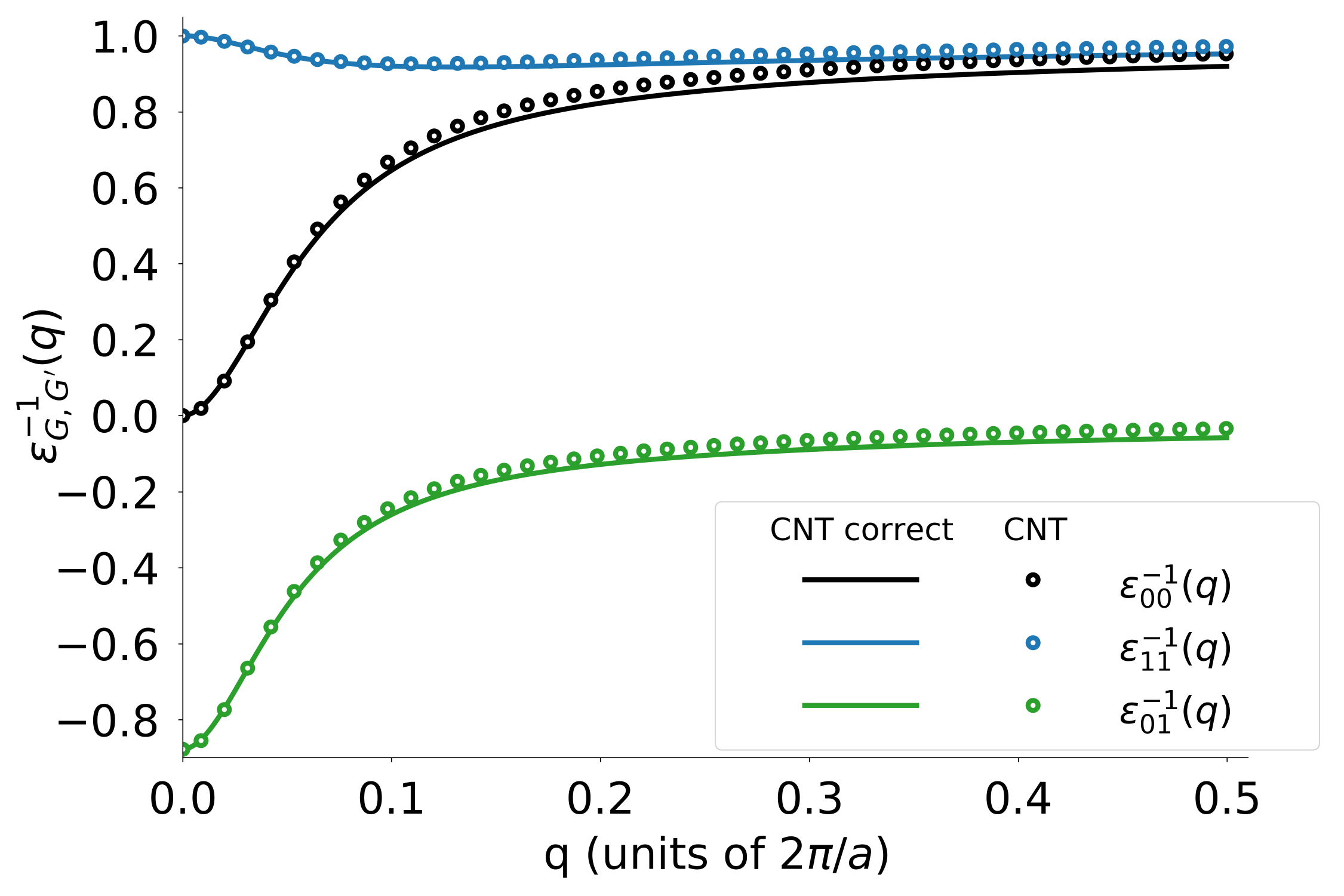}
\centering
\caption{
Corrected vs original matrix elements of the model dielectric function for the armchair (3,3) nanotube. Head term, $\epsilon^{-1}_{0,0}(\boldsymbol{q})$, diagonal term, $\epsilon^{-1}_{\boldsymbol{G}_{\perp},\boldsymbol{G}_{\perp}}(\boldsymbol{q})$, and wing term, $\epsilon^{-1}_{0,\boldsymbol{G}_{\perp}}(\boldsymbol{q})$, vs $q$ for the smallest vector $\boldsymbol{G}_{\perp}$ with $n_1=1$, $n_2=n_3=0$. 
\label{eps33-comp}}
\end{figure}

\begin{figure}
\includegraphics[width=8.6cm,trim={0.5cm 0.5cm  0.0cm 0.4cm},clip]{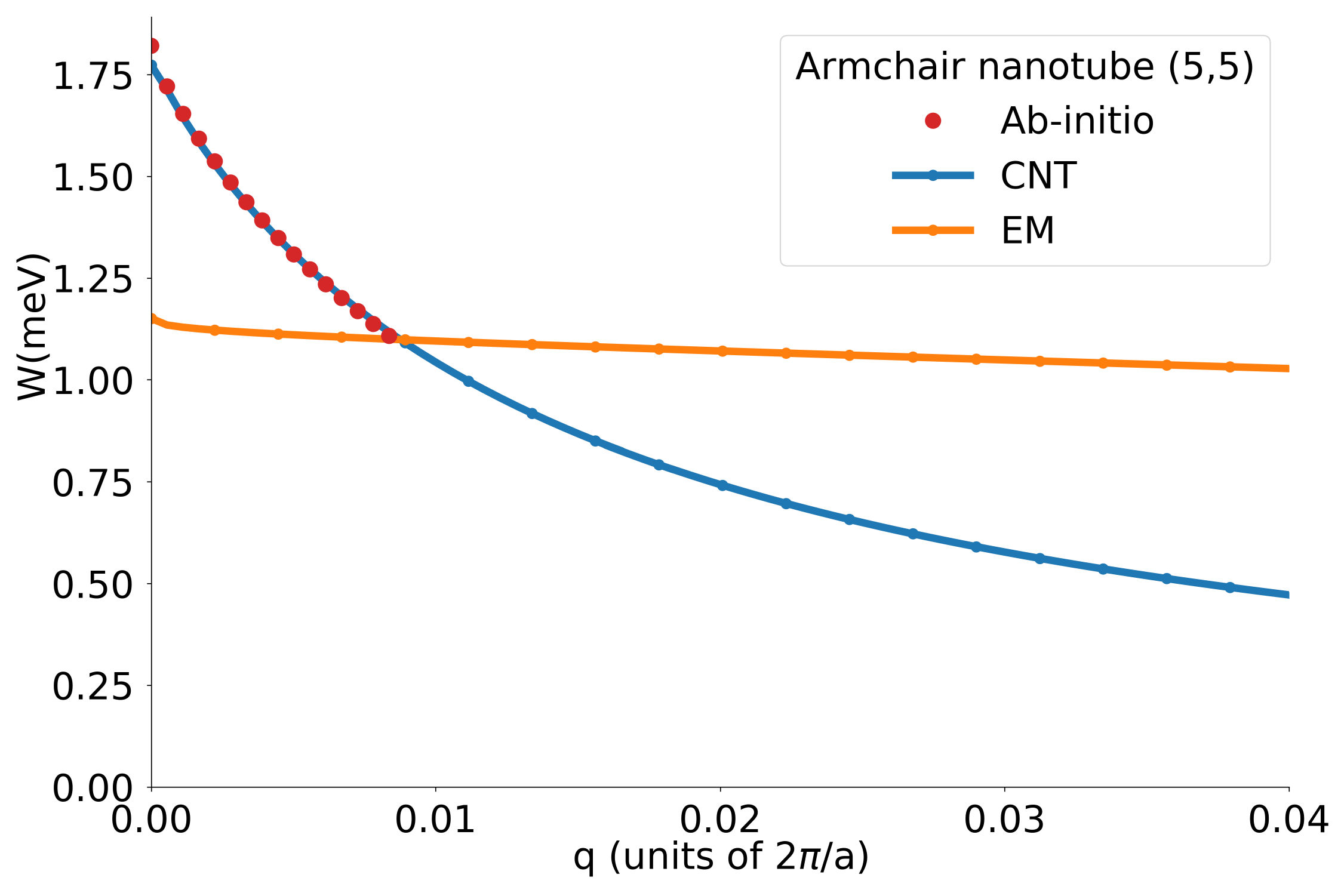}
\centering
\caption{
Screened electron-hole interaction $W(q)$ vs $q$ for the armchair (5,5) nanotube derived from first-principles (ab-initio),  effective mass (EM) and  and two-band model (CNT) approaches. 
\label{potarmchair-2}}
\end{figure}

In order to achieve high numerical accuracy, we correct the model polarization $\Pi$ through a multiplicative factor,
$q$- and $\theta$-dependent, which very slightly differs from unity. We derive this factor by fitting the macroscopic, first diagonal, and first wing terms of  $\Pi$
to first-principles data, according to 
\begin{multline}
\Pi^{\text{correct}}_{\boldsymbol{G},\boldsymbol{G'}}(\boldsymbol{q}) \quad = \quad  \Pi^{\text{CNT}}_{\boldsymbol{G},\boldsymbol{G'}}(\boldsymbol{q})  \\ \times\quad \left\{5 \cos \! \left[2.7\left(\pi/6- \theta\right) \! \right] \! R q + 3.806\, [R/(\text{1 nm})]^{1.46}  \right\}.
\end{multline}
Here the numerical coefficients fit the first-principles matrix elements of armchair  tubes (3,3), (4,4), (5,5), and zigzag tubes (9,0), (12,0). As this correction is immaterial for zigzag tubes, we employ the corrected form $\Pi^{\text{correct}}$ throughout the paper.
For the sake of illustration, we compare the corrected and uncorrected terms of $\epsilon^{-1}$ for the (3,3) armchair tube in Fig.~\ref{eps33-comp}, the discrepancies being small and only relevant at short wavelength. The correction of $\Pi$ allows for an excellent matching between model and first-principles predictions of the dressed electron-hole interaction in armchair tubes, as shown in Fig.~\ref{potarmchair-2}.


Figure \ref{potarmchair-2} illustrates the key findings of this paper. 
The EM theory (orange curve) predicts that the dressed electron-hole attraction depends weakly on the transferred momentum $q$ in gapless tubes, 
hence corresponding to
a short-range force. The force range is given by the expression \eqref{eq:epsilonEM} of $\Pi_{\text{EM}}$ for $k_{\tau}\rightarrow 0$, i.e.,
$\Pi_{\text{EM}}(q)=-4A/\pi\gamma$, with $4A/\pi\gamma$ being the density of states. This is just the Thomas-Fermi result for an effectively one-dimensional metal. On the contrary, both first-principles (red dots) and two-band-model calculations (blue curve) predict that $W$ has a singular-like profile at long wavelength---roughly logarithmic~\cite{varsano2017carbon}---signaling that the force binding electrons and holes is actually long-ranged. This is a substantial effect of microscopic local fields, which emerges as electrons effectively move on a cylindrical surface and not on a line. As a consequence, gapless tubes are unstable against the spontaneous condensation of excitons,\cite{varsano2017carbon} whereas the EM theory\cite{Ando1997}
predicts the exciton binding energy to vanish with the gap.

\begin{table}[]
    \centering
    \begin{tabularx}{0.98\columnwidth}{ XXX }
    \hline
      &  Triplet &  Singlet\\
     \hline
    \rowcolor{gray!30}
    Ab initio (Ref.~27) &   -7.91meV & -6.10 meV \\

    CNT correct   & -7.07 meV  & -5.22 meV \\
    
    \rowcolor{gray!30}
    CNT, tiny gap &  -5.79meV   & -4.87meV \\

    CNT, gapless & -2.00 meV & -1.13 meV \\
      \hline
    
    \end{tabularx}
    \caption{Excitation energy of the lowest lying triplet and singlet exciton in the (3,3) armchair carbon nanotube from first-principles (ab intio) and two-band-model (CNT) approaches.}
    \label{tab:exciton-energy}
\end{table}

We will use the results of the present work to treat excitonic effects in narrow-gap NTs elsewhere. In order to complete our discussion of gapless tubes, here we reconsider the calculation of exciton properties from first principles reported in Ref.~\onlinecite{varsano2017carbon}. In the calculation by Varsano {\it et al.}\cite{varsano2017carbon} for the (3,3) armchair tube, the system was actually gapped by a tiny quantity, 1.08 meV, arising from the numerical discretization of the reciprocal space.  In the following we show that this artefact does not harm the claim of excitonic instability. 

First, the tiny gap does not affect the calculation of $\epsilon^{-1}$ reported in Figs.~\ref{FIG.epss}(b) and \ref{eps-micro}(b) in any way, since: (i) the reciprocal-space mesh in energy units,  $\gamma\, \text{d}q = 1.5$ meV, is obviously larger than the gap (ii) the computed macroscopic inverse dielectric constant, $\epsilon^{-1}_{0,0}(q)$, already vanishes at small momenta $q > \text{d}q$, as apparent from Fig.~\ref{FIG.epss}(b).

Furthermore, we checked the effect of the artificial gap on the exciton energy, by numerically solving the Bethe-Salpeter equation within the two-band model for the screened interaction presented in this paper. The resulting excitation energies of the lowest singlet and triplet excitons are reported in Table \ref{tab:exciton-energy} (CNT correct) for the gap being exactly zero, and compared with the first-principles results (ab initio). The discrepancies are minor, smaller than 1 meV and of the order of magnitude of the artificial gap. One might also wonder whether the two-band-model calculation performed without applying the corrective factor to $\Pi$, which fits first-principles data, were still able to predict the excitonic instability. The results of such calculation, respectively in the presence of the tiny gap (CNT, tiny gap) and in the gapless case (CNT, gapless), are reported in the last two rows of Table \ref{tab:exciton-energy}. In all events the excitation energy of the exciton remains negative, which points to the tendency of excitons to spontaneously form.

\begin{figure*}[t!]

\begin{tabular}{cccc}
\hspace*{-3mm}
\textbf{a} & & \hspace*{0.3cm} \textbf{b} & \\
 & \hspace*{-0.7cm} \includegraphics[width=8.6cm,trim={1.0cm 0.5cm 2.5cm 2.5cm},clip]{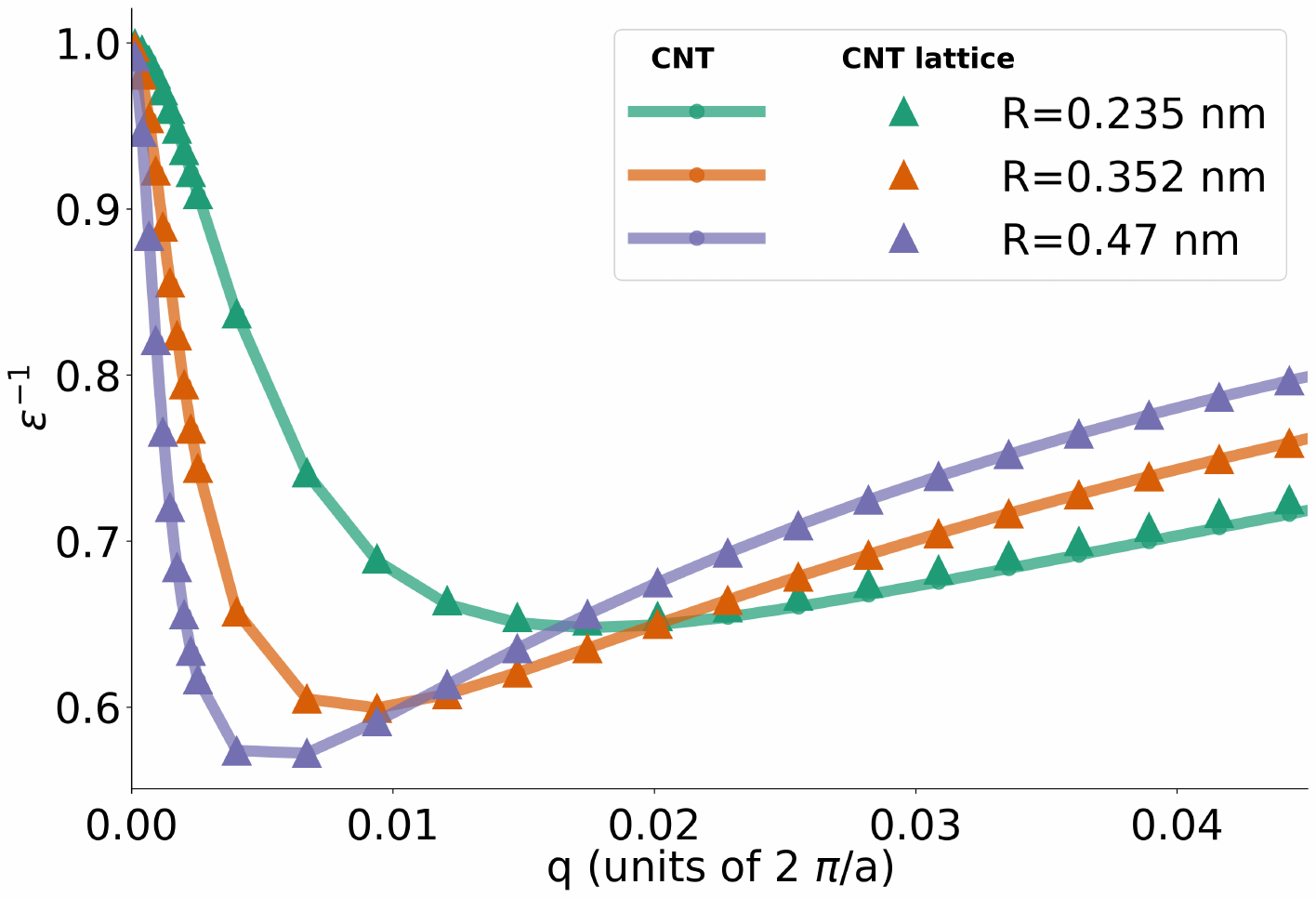}
& & \hspace*{-0.7cm}
\includegraphics[width=8.6cm,trim={1.0cm 0.5cm 2.5cm 2.5cm},clip]{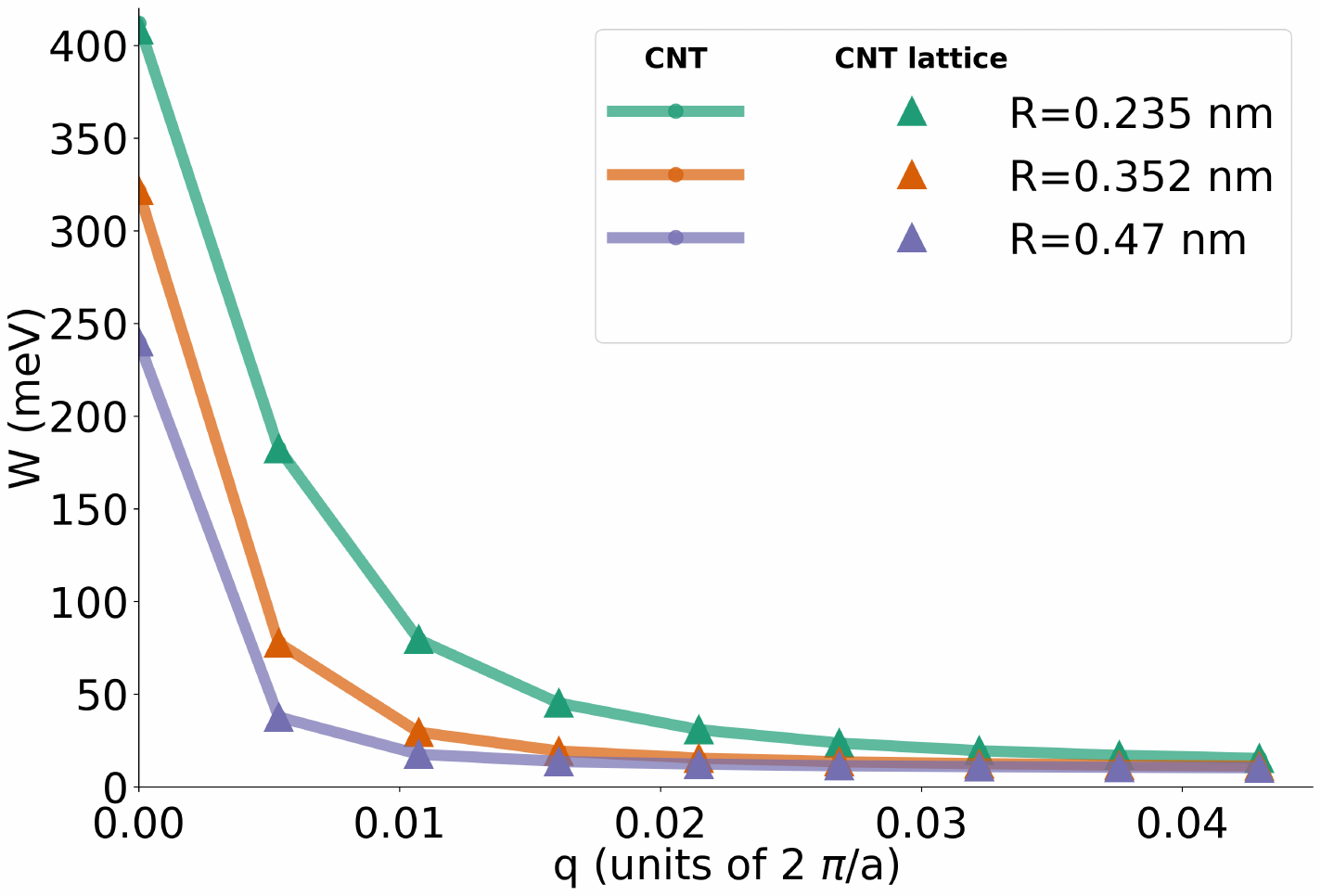}\\

\textbf{c} & & \hspace*{0.3cm} \textbf{d}\\
& \hspace*{-0.7cm} \includegraphics[width=8.6cm,trim={1.0cm 0.5cm 2.5cm 2.5cm},clip]{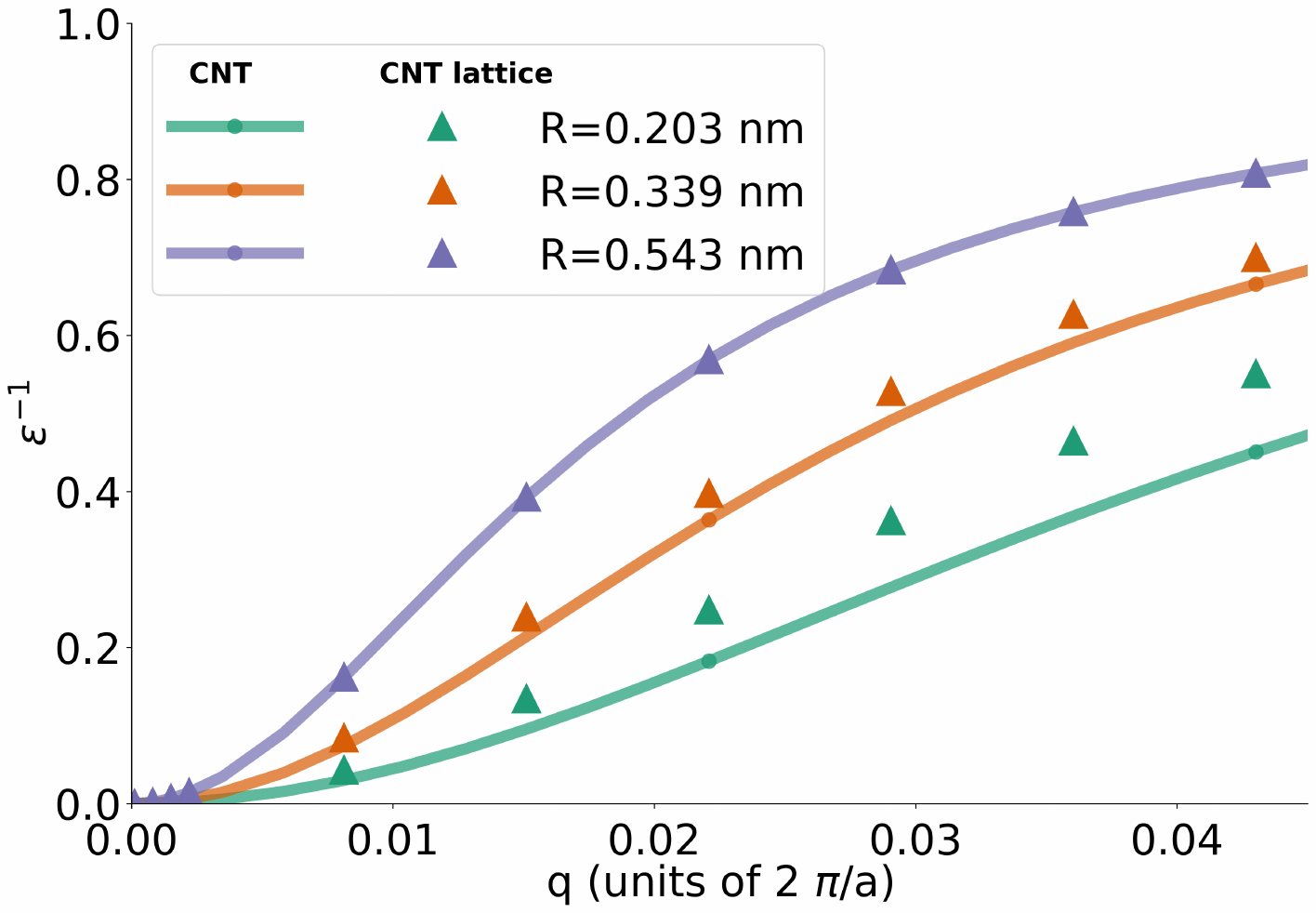} & & \hspace*{-0.7cm}
\includegraphics[width=8.6cm,trim={1.0cm 0.5cm 2.5cm 2.5cm},clip]{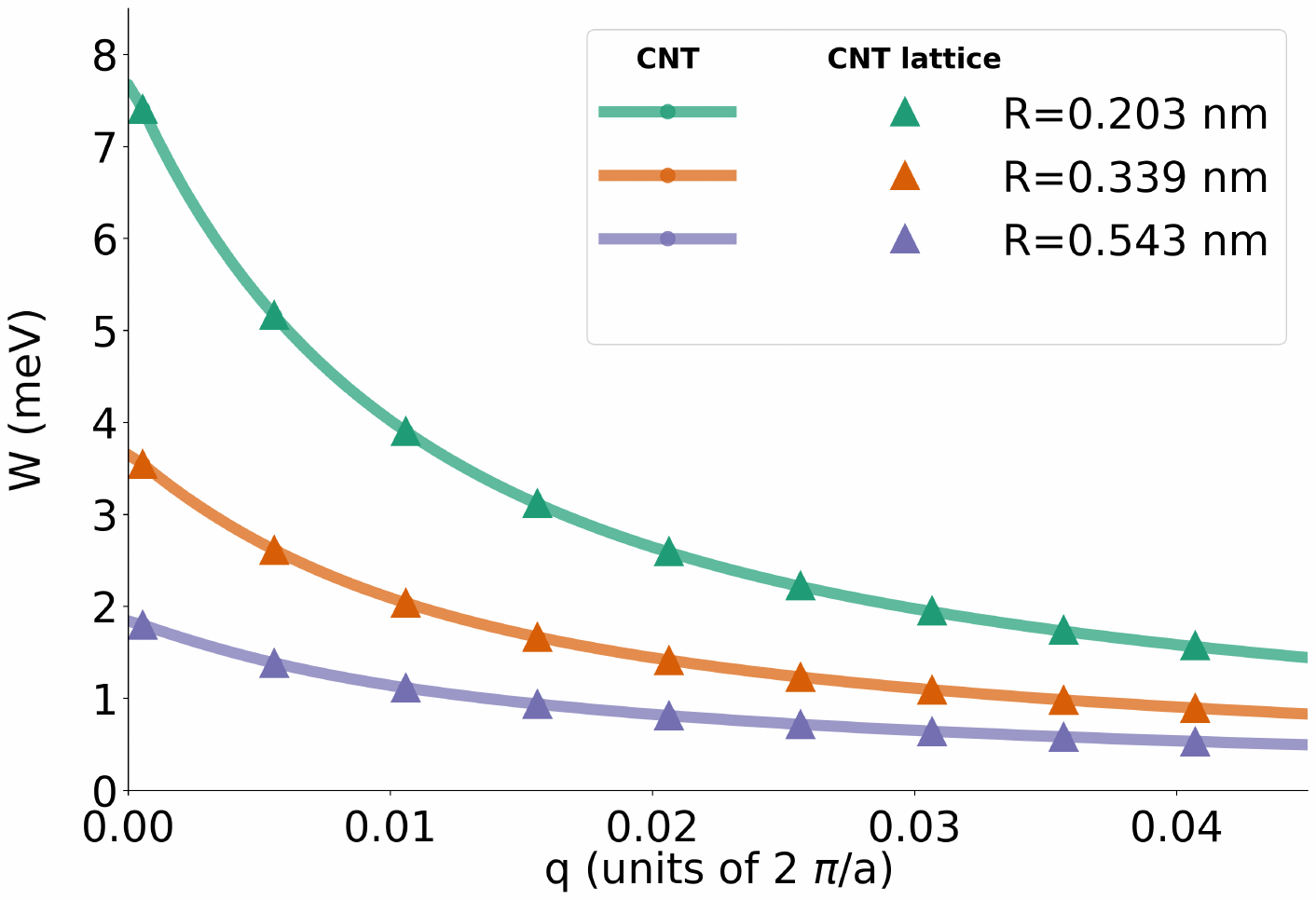}
\end{tabular}
\caption{\label{FIG:hom-ex} 
Macroscopic inverse dielectric function, $\epsilon_{0,0}^{-1}(q)$ (panel 
a and c),
and screened electron-hole interaction, $W(q)$ (panel b and d), vs momentum, $q$,
computed from different model approaches for selected zigzag (a and b)
and armchair (c and d) nanotubes, respectively. Data are derived by considering either the lattice structure  (triangles) or by spreading homogeneously the electronic charge (solid curves) over the cylindrical tube surface. The selected zigzag tubes in panel a and b have chiral indeces (6,0), (9,0), and (12,0). The armchair tubes, in panel c and d, have chiral indeces (3,3), (5,5), and (8,8). }
\end{figure*}

\subsection{Validation of the structural model through comparison with results for armchair and zigzag lattices}\label{sec:jellium}

Throughout this work we model the carbon nanotube structure as a cylindrical surface over which the electrons occupying the Bloch states $\psi(\boldsymbol{r})$, which multiply the envelopes in Eq. \eqref{eq.Bloch}, are spread homogeneously according to the ansatz of Eq.~\eqref{eq:Blochjellium} (see Subsec.~\ref{ssec:3d}).
In this subsection we validate this model by comparing both the dielectric function and the screened electron-hole interaction with those computed by considering the actual location of atoms in the curved honeycomb lattice.
To this aim, we replace the ``jellium'' of Eq.~\eqref{eq:Blochjellium} 
with orbitals localized on either the zigzag or the armchair lattice,
as detailed in Appendixes \ref{zigzag-ex-ov} and \ref{armchair-ex-ov}, respectively. This change affects the overlap integrals that enter the expressions of the dielectric function (Eq.~\ref{epscnt}) and screened electron-hole interaction (Eq.~\ref{eq:Wtcnt}). As we show below, the discrepancies are minor.

\subsubsection{Zigzag lattice}

Zigzag carbon nanotubes $(n,0)$ have a chiral vector $\boldsymbol{\mathcal{C}}= n \boldsymbol{a} $ stretching over $n$ units cells of graphene (the vectors $\boldsymbol{a}$ and $\boldsymbol{b}$ are shown in  Fig.~\ref{FIG:graphsheet}). Whereas in the simpler structural model the $n$ units cells are represented as two rings and the electron charges are spread into a ``jellium'', here we consider all $n$ cells and all $2n$ atom positions per sublattice. The overlap integrals, derived in Appendix \ref{zigzag-ex-ov}, are:
\begin{multline}
\label{eq.inter_}
\langle \alpha k|  e^{-i (\boldsymbol{G} + \boldsymbol{q}) \cdot \boldsymbol{r}} | \alpha' k + q  \rangle = \\  \frac{1}{2}  \left[  \! \! \frac{k (k+q) +k_{\tau}^2-i q k_{\tau}}{\sqrt{k^2+k_{\tau}^2}\sqrt{(k+q)^2+k_{\tau}^2}} + (2 \delta_{\alpha,\alpha'} -1)  \! \right]  \\ \times \quad \left[J_0(R G_{\perp}) + 2 (-1)^n  J_{2n} (R G_{\perp})\right],
\end{multline}
with $J_{2n} $ being the Bessel function of first kind of order $2n$. This overlap integral is similar to the ``jellium'' expression \eqref{eq.ovecnt} except for the correction due to the higher order Bessel function, the order being linked to the number of atoms in the cell.

Similarly, the dielectric function is:
\begin{align}
\label{eq.epszz} 
\epsilon^{\text{zigzag}}_{\boldsymbol{G},\boldsymbol{G'}}(\boldsymbol{q})  = \notag \quad & \delta_{\boldsymbol{G},\boldsymbol{G'}} + \frac{2 A}{\pi \gamma} \  v(\boldsymbol{q} + \boldsymbol{G}) \\ \notag
\times \quad & \left[J_0(R G_{\perp})  + 2 (-1)^n     J_{2n} (R G_{\perp})\right]  
\\ \notag \times \quad &
\left[J_0(R G_{\perp}') + 2 (-1)^n  J_{2n} (R G_{\perp}')\right]   \\\times \quad & \sum_\tau \Bigg[ 1 
 +   \frac{ 2 k_{\tau}^2}{q\sqrt{q^2 + 4 k_{\tau}^2}} \log \Bigg(\frac{\sqrt{q^2 + 4 k_{\tau}^2} - q}{\sqrt{q^2 + 4 k_{\tau}^2} + q}\Bigg) \Bigg].
\end{align}

The inverse macroscopic dielectric function derived above, $[\epsilon_{0,0}^{\text{zigzag}}]^{-1}(q)$, as well as the screened electron-hole interaction, $W^{\text{zigzag}}(q)$, are reported for selected zigzag tubes in Figs.~\ref{FIG:hom-ex}(a) and (b), respectively (triangles, CNT lattice).
The results are essentially identical to those derived from the simpler structural model used throughout the paper (solid curves, CNT).

\subsubsection{Armchair lattice}

Armchair carbon nanotubes $(n,n)$ have a chiral vector $\boldsymbol{\mathcal{C}}= 2n \boldsymbol{a} + n \boldsymbol{b}$ corresponding to  the chiral angle $\theta=\pi/6$. The vector $\boldsymbol{\mathcal{C}}$ extends over $2n$ units cells of graphene. The overlap integrals of armchair nanotubes, derived in Appendix \ref{armchair-ex-ov}, take into account the locations of the atoms occupying these $2n$ units cells:
\begin{multline}
 \!  \!  \! \langle \alpha k|  e^{-i (\boldsymbol{G} + \boldsymbol{q}) \cdot \boldsymbol{r}} | \alpha' k + q  \rangle =  \\  \!  \!  \! \frac{1}{2}  \left\{  
\text{sign}[k (k+q)] 
\,+ (2 \delta_{\alpha,\alpha'} -1)  \! \right\}   J_0( \! R G_{\perp}  \! )\quad  + \\       \!  \!  \! \frac{(-1)^n}{2} \! \Bigg[   2\,
\text{sign}[k (k+q)] 
- ( 2 \delta_{\alpha,\alpha'} -1)  \! \Bigg]   J_{2n}(  \! R G_{\perp}  \! ).
\end{multline}
This overlap integral differs from the ``jellium'' expression \eqref{eq.ovecnt}  in the  addition of an extra term, originating by the Bessel function of order equal to the number of unit cells. This in turn changes the dielectric function, through the occurrence of an extra, cut-off dependent term:
\label{eq.epsac}
\begin{multline}
\epsilon^{\text{armchair}}_{\boldsymbol{G},\boldsymbol{G'}}(\boldsymbol{q})= \delta_{\boldsymbol{G},\boldsymbol{G'}} + \frac{A}{\pi \gamma}  v(\boldsymbol{q} + \boldsymbol{G}) \Bigg[(2\, J_0(R G_{\perp}) \quad + \\ (-1)^n J_{2n}(R G_{\perp})) \, (2 J_0(R G'_{\perp}) + (-1)^n J_{2n}(R G'_{\perp})) \quad + \\
\frac{9}{2}  J_{2n}(R G_{\perp}) J_{2n}(R G'_{\perp}) \log \left( \frac{4 k_o^2}{q^2} - 1\right) \Bigg].
\end{multline}
The extra-term ensures that the dielectric function diverges for $q\rightarrow 0$, the expected behaviour in gapless tubes. 

The inverse macroscopic dielectric function derived above, $[\epsilon_{0,0}^{\text{armchair}}]^{-1}(q)$ [triangles, CNT lattice in Fig.~\ref{FIG:hom-ex}(c)]
differs only slightly from that derived from the simpler structural model (solid curves, CNT), and only for $q>0.01 (2\pi)/a$ and small radii. Importantly, these small discrepancies are irrelevant for the computation of the screened electron-hole interaction, as apparent from Fig.~\ref{FIG:hom-ex}(d).

\subsection{Super Coulombic interaction}\label{sec:super} 

\begin{figure}[t!]
\hspace*{-3mm}
\includegraphics[width=8.6cm,trim={2.0cm 0.5cm 2.5cm 2.5cm},clip]{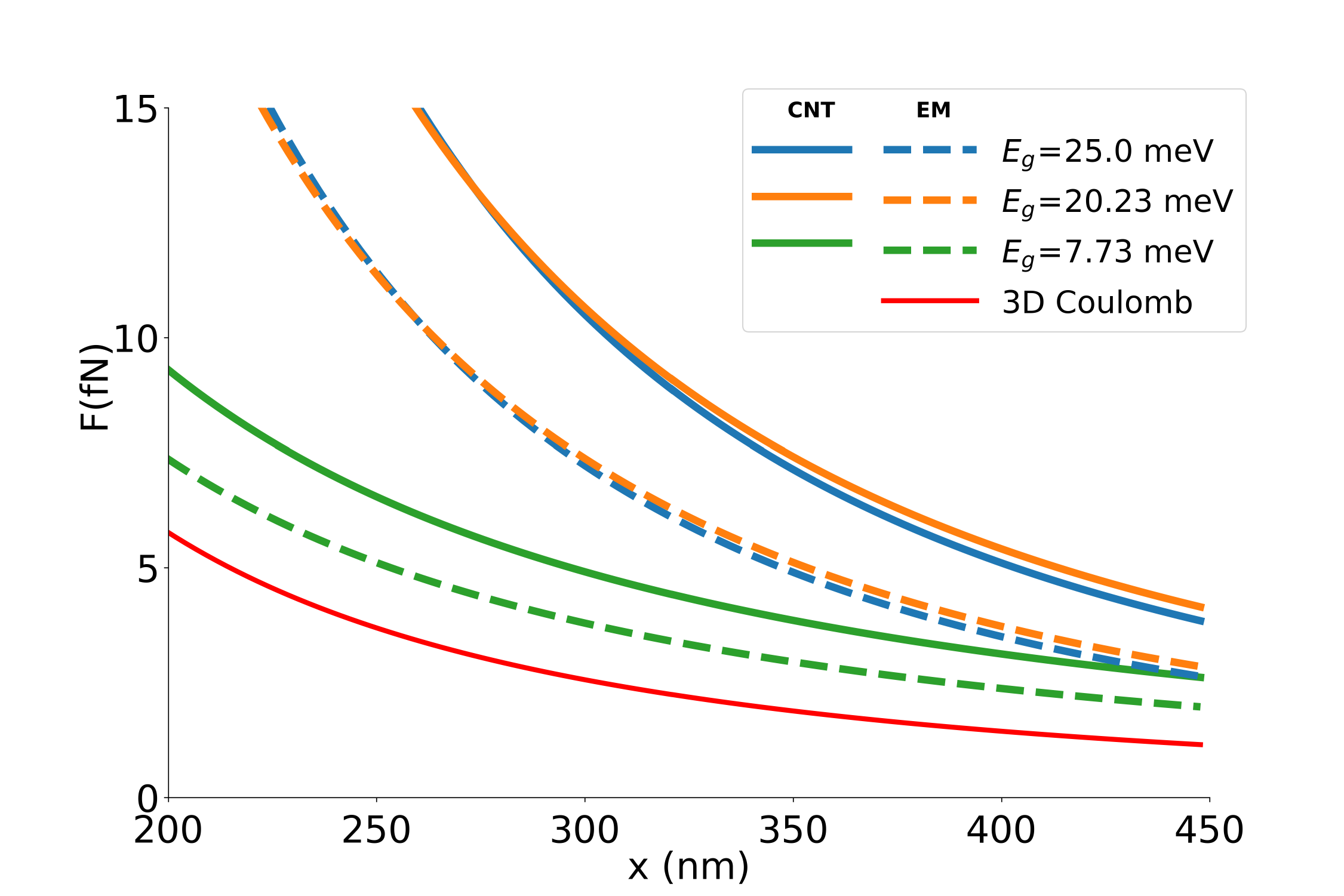}
\centering
\caption{Effective electron-electron force along the nanotube axis vs electron separation, $x$, in tubes having different energy gaps, $E_g$. The tube radius is $R=1$ nm. The solid and dashed curves are respectively the two-band-model calculation (CNT) and the effective-mass (EM) prediction. The red curve is the standard three-dimensional Coulomb force.\label{R=1_2}}
\end{figure}

\begin{figure}[t!]
\hspace*{-3mm}
\includegraphics[width=8.6cm,trim={2.0cm 0.5cm 2.5cm 2.5cm},clip]{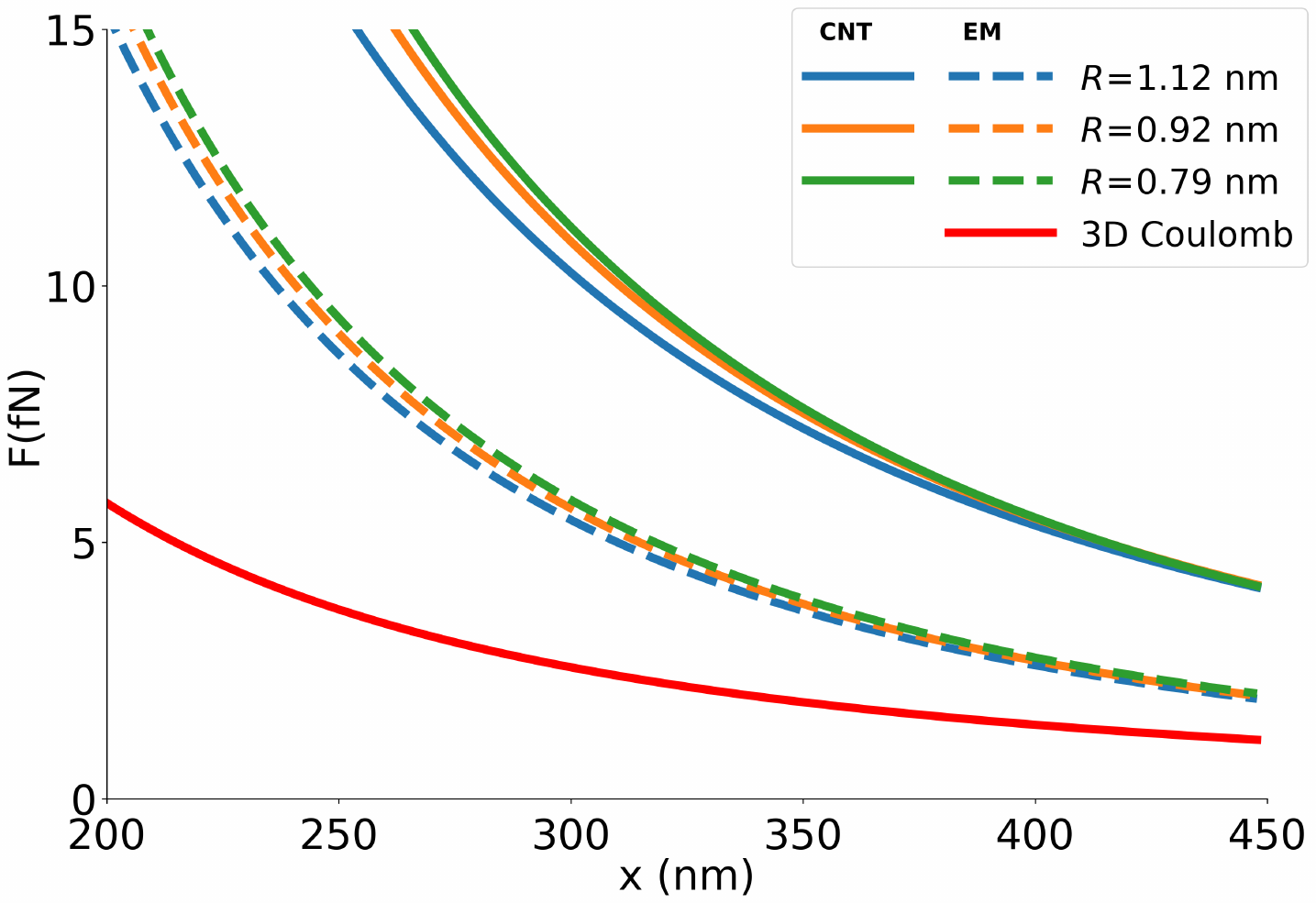}
\centering
\caption{
Effective electron-electron force along the nanotube axis vs electron separation, $x$, in tubes of different radii, $R$. The energy gap is $E_g$= 20 meV.  The solid and dashed curves are respectively the two-band-model calculation (CNT) and the effective-mass (EM) prediction. The red curve is the standard three-dimensional Coulomb force.
\label{Eg=20_2}}	
\end{figure}

Direct measurements of electron-electron interaction in materials are generally hard to perform, due to the interference between the measured system and the probe. Recently, the group of S.~Ilani at Weizmann Institute of Science developed a new sensing technique to minimize such interference,\cite{Ilanit2019} by means of using a suspended carbon nanotube as a scanning tool to probe, with minimal invasiveness, few-electron states within another nanotube. A new experiment, which focused on the case of just two electrons populating the $c$ band of a narrow-gap nanotube, was able to directly
measure the Coulomb force repelling the two charges in real space.\cite{shapireeinteraction} 

Therefore, we have computed the screened electron-electron interaction, projected onto the $c$ band and Fourier-transformed in real space. Figures \ref{R=1_2} and \ref{Eg=20_2} show the force dependence on the electron separation in the range of hundreds of nm, which is relevant to the experiment of Ref.~\onlinecite{shapireeinteraction}, and compare it to the standard, three-dimensional bare Coulomb force (red curve). Both effective-mass (EM) and two-band-model (CNT) calculations predict that the effective force is stronger than the bare Coulomb force, regardless of the gap (Fig.~\ref{R=1_2}) or radius (Fig.~\ref{Eg=20_2}) size, as a consequence of the non-local character of screening in nanotubes.\cite{Deslippe2009} However,
only the inclusion into the model of microscopic local fields, induced by the motion of electrons on the curved tube surface, leads to a major enhancement of the effective force, as seen by contrasting CNT and EM curves for given tube.

\section{Conclusions}\label{sec:conclusions}

In this work we have developed a simplified approach to compute the dielectric function of narrow-gap carbon nanotubes of any size and chirality, which has the same accuracy of first-principles calculations but is computationally cheaper. A detailed analysis shows a giant 
enhancement of the screened Coulomb interaction at long wavelength with respect to the one expected for an effectively one-dimensional system. We find that this is caused by the strong, microscopic local fields generated by the electron motion on the curved tube surface. The paradoxical consequence is that the screened electron-hole interaction, once projected onto the lowest conduction and highest valence band, remains long-ranged even in the presence of Fermi points (armchair tubes). 

Our findings provide a deeper insight into our previous claim\cite{varsano2017carbon} that undoped narrow-gap carbon nanotubes are excitonic and not Mott insulators. Furthermore, the calculated electron-electron interaction in real space shows that the force is super Coulombic beyond expectations. We anticipate our theory lays the quantitative basis for future studies of many-body physics in carbon nanotubes, where the long-range character of interaction leads to novel phenomena. 

\section*{Acknowledgements}

We thank Catalin Spataru and Shahal Ilani for many discussions over the years concerning many-body physics in carbon nanotubes. 

This work was supported in part by the “MAterials
design at the eXascale” (MaX) European Center of Excellence (www.maxcentre.eu) funded by the European Union H2020-INFRAEDI-2018-1
program, Grant 824143. It was also supported by the Italian national
program PRIN2017 2017BZPKSZ “Excitonic insulator in two-dimensional
long-range interacting systems.” We acknowledge access to the
Marconi supercomputing system based at CINECA, Italy, through Partnership for Advanced
Computing in Europe (PRACE) and the Italian SuperComputing Resource Allocation
program (Iscra).

\bibliography{./biblio/cnt-biblio,./biblio/ei}

\appendix

\section{Overlap integrals within the model of the nanotube as a series of rings}
\label{appendix}

In this Appendix we compute the overlap integrals, $\rho_{cv}$, between conduction and valence band states, by modelling the carbon nanotube as a series of rings over which the electronic charge is spread   homogeneously. 

We recall that the orbital wave function of NT states is:
\begin{equation}
\label{eq:defff}
\Psi_{\alpha \tau   k}(\boldsymbol{r})= F^{\tau A}_{\alpha k  }(\boldsymbol{r}) \,\psi_{\tau A} (\boldsymbol{r}) + F^{\tau B}_{\alpha k  }(\boldsymbol{r})\, \psi_{\tau B} (\boldsymbol{r}).
\end{equation}
Here the envelope functions, $F$, take the form \eqref{eq.envelope}, and the Bloch states, $\psi_{\tau \eta} (\boldsymbol{r})$, are spread homogeneously over rings placed along the $y$ axis:
\begin{align}
\psi_{\tau A} (\boldsymbol{r})= \frac{1}{\sqrt{N}} \frac{e^{i \phi_{\tau A}}}{\sqrt{2 \pi R }}\sum_{l=1}^{N} \left[ e^{i \boldsymbol{K_\tau} \cdot \boldsymbol{R}^A_{l}} g(\boldsymbol{r}-\boldsymbol{R}^A_{l}) \right], \\
\psi_{\tau B} (\boldsymbol{r})= \frac{1}{\sqrt{N}} \frac{e^{i \phi_{\tau B}}}{\sqrt{2 \pi R}} \sum_{l=1}^{N} \left[ e^{i \boldsymbol{K_\tau} \cdot \boldsymbol{R}^B_{l}} g (\boldsymbol{r} -\boldsymbol{R}^B_{l}) \right],
\end{align}
the phases being $\phi_{K A}=0$, $\phi_{K' A}=\theta$, $\phi_{K B}=-\frac{\pi}{3}+\theta$ and $\phi_{K B}=0$. 
The functions $g$ are localised along the $y$ axis and orthogonal, according to:
\begin{eqnarray}
g^*\!(\boldsymbol{r}-\boldsymbol{R}^{\eta}_{l})\,g(\boldsymbol{r}-\boldsymbol{R}^{\eta'}_{l'}) & = & \delta_{\eta,\eta'}  \delta_{l,l'}  \nonumber \\ & \times & \delta(\rho-R) \,\delta(y-{R}^{\eta}_{l}).
\end{eqnarray}
There are $N$ charge rings for given sublattice and the rings within the sublattice are uniformly spaced along the NT axis, at distance $\lambda = a \cos \left(\frac{\pi}{6}-\theta \right)$:
\begin{align}
\Bigg\{ \begin{array}{c}
\boldsymbol{R}^A_{l}= \left(l \lambda + y^A_0\right) \hat{y} \ \ \\
\boldsymbol{R}^B_{l}= 
\left( l \lambda + y^B_0\right)
\hat{y} \ \   
\end{array} \ \ \ \textrm{with} \ \ \ l = 0,...,N-1,
\end{align}
with $y^A_0=y^B_0=0$. 

The overlap integrals of interest within each valley $\tau$, $\rho_{cv}$, are:
\begin{align}
\rho_{\alpha\alpha'}(k\hat{y},q\hat{y},\boldsymbol{G})=
\langle \alpha \tau k| e^{-i (\boldsymbol{G} + \boldsymbol{q}) \cdot \boldsymbol{r}}   | \alpha'  \tau k + q  \rangle,
\end{align}
where the ket $ | \alpha \tau k \rangle$ indicates the NT orbital wave function $\Psi_{\alpha \tau   k}$. To proceed we make the expression \eqref{eq:defff} explict and insert it into the definition of $\rho_{\alpha\alpha'}$, using cylindrical coordinates as well as the orthogonality of $g$'s :
\begin{align}
& \rho_{\alpha\alpha'} = \langle \alpha \tau k| e^{-i (\boldsymbol{G} + \boldsymbol{q}) \cdot \boldsymbol{r}}   | \alpha'  \tau k + q  \rangle = \\ 
& \frac{1}{2 \pi R N} \sum_{\eta=A}^B (F^{ \eta}_{\tau \alpha k})^{*}  F^{\eta}_{\tau \alpha' k+q} \int_{0}^{\infty} \! \! \!  \! d\rho \int \! \! d  y \int_{0}^{2 \pi} \! \! \!  \! d \varphi \\& \times \quad 
\sum_{l=0}^{N-1} \rho \ \delta \left( \rho -R \right) \notag  \delta \left( y -l \lambda \right) e^{-i \left( G_{\parallel}  y + \rho G_{\perp}  \cos{\varphi} \right)}.
\end{align}
The delta function of  $\rho$ argument constrains integration over the nanotube surface:
\begin{align}
& \rho_{\alpha\alpha'} =
\frac{1}{2 \pi N} \sum_{\eta=A}^B (F^{ \eta}_{\tau \alpha k})^{*}  F^{\eta}_{\tau \alpha' k+q} \int \! \! d  y \int_{0}^{2 \pi} \! \! \!  \! d \varphi  
\\& \times \quad
\sum_{l=0}^{N-1}  \notag   \delta \left( y -l \lambda \right)  e^{-i \left( G_{\parallel}  y + R G_{\perp}  \cos{\varphi} \right)}. 
\end{align}
We then integrate over $y$ 
and recall the expression \eqref{eq.G} of $G_{\parallel}$, obtaining:
\begin{align}
& \rho_{\alpha\alpha'} =
 \frac{1}{2 \pi N}  \sum_{\eta=A}^B (F^{ \eta}_{\tau \alpha k})^{*}  F^{\eta}_{\tau \alpha' k+q}\\&\times \quad   \sum_{l=0}^{N-1}   e^{-i 2 \pi n_2 l} \! \int_{0}^{2 \pi} \! \! \!  \! d \varphi \ e^{-i R G_{\perp} \! \cos(\varphi) }.
\end{align}
The indices $n_2$ and $l$ are both integers, then the exponential is always equal to $1$. Furthermore, 
the integral over $\varphi$ is equal to $2 \pi J_0(R G_{\perp})$. In conclusion, the overlap integral is:
\begin{multline} 
\langle \alpha k|  e^{-i (\boldsymbol{G} + \boldsymbol{q}) \cdot \boldsymbol{r}} | \alpha' k + q  \rangle \quad= \\ \left[ (F^{A}_{\tau  \alpha k})^{*} F^{A}_{\tau  \alpha' k+q} + (F^{B}_{\tau  \alpha k})^{*} F^{B}_{\tau \alpha' k+q}) \right] J_0(R G_{\perp}). 
\end{multline}

The possible combinations of $\alpha$ and $\alpha'$ are as follows. The intraband overlap integrals have either $\alpha=c$ and $\alpha'=c$ or  $\alpha=v$ and $\alpha'=v$: 
\begin{multline}
\label{eq.inter} 
\langle c k|  e^{-i (\boldsymbol{G} + \boldsymbol{q}) \cdot \boldsymbol{r}} | c k + q  \rangle = \langle v k|  e^{-i (\boldsymbol{G} + \boldsymbol{q}) \cdot \boldsymbol{r}} | v k + q  \rangle = \\  \frac{1}{2}  \left(  \! \! \frac{k (k+q) +k_{\tau}^2-i q k_{\tau}}{\sqrt{k^2+k_{\tau}^2}\sqrt{(k+q)^2+k_{\tau}^2}} + 1  \! \right)  J_0(R G_{\perp}). 
\end{multline}
The interband overlap integrals, with $\alpha=c$ and $\alpha'=v$ or viceversa, exhibit a negative sign instead:
\begin{multline}
\label{eq.intra} 
\langle c k|  e^{-i (\boldsymbol{G} + \boldsymbol{q}) \cdot \boldsymbol{r}} | v k + q  \rangle = \\  \frac{1}{2} \left(  \! \! \frac{k (k+q) +k_{\tau}^2-i q k_{\tau}}{\sqrt{k^2+k_{\tau}^2}\sqrt{(k+q)^2+k_{\tau}^2}} - 1  \! \right) 
 J_0(R G_{\perp}).
 \end{multline}

\section{Overlap integrals of zigzag carbon nanotubes}
\label{zigzag-ex-ov}

In this Appendix and in the next one we compute the overlap integrals $\rho_{cv}$ for zigzag and armchair carbon nanotubes, respectively, by accounting for the actual location of atoms on the curved honeycomb lattice. To this aim, we replace the Bloch states
\eqref{eq:Blochjellium}
with orbitals localised along the tube circumference, whose positions depend on the azimuthal angle $\varphi$. 
The derivation of $\rho_{cv}$ proceeds analogously to what done in Appendix \ref{appendix} until one performs the integration  over $\varphi$. As $n$ atoms per sublattice are now localised along $\varphi$,  one has to sum over their positions, the overlap integrals being:
\begin{align}
\label{eq.int-phi2} 
& \rho_{\alpha\alpha'} =
 \frac{1}{2 n}  \sum_{\eta=A}^B (F^{ \eta}_{\tau \alpha k})^{*}  (F^{\eta}_{\tau \alpha' k+q})\nonumber
 \\
 &\times\quad\int_{0}^{2 \pi} \! \! \!  \! d \varphi \sum^{n-1}_{j=0} \delta \left( \varphi- \varphi^{\eta}_{j} \right) e^{-i R G_{\perp} \! \cos(\varphi)} ,
\end{align}
where the $\varphi^{\eta}_{j}$ are the angular coordinates of the atoms. We specialise to zigzag nanotubes in the following.
    
Zigzag nanotubes $(n,0)$ may be thought of as built by rolling the graphene sheet along the chiral vector $\boldsymbol{\mathcal{C}}= n \boldsymbol{a}$, which stretches over $n$ units cells of graphene.
There are $2n$ atoms per sublattice, whose positions on the tube surface are:
\begin{align}
\label{zz-coor}
\boldsymbol{R}^A_{l,j}=& 
\Bigg\{ \begin{array}{c}
  R \ \frac{2 \pi j}{n} \hat{\varphi} + l \lambda \hat{y} + + R \hat{\rho}  \\ 
  R \ \frac{\pi (2 j + 1)}{n} \hat{\varphi} + \left( l \lambda + \frac{
\sqrt{3}}{2} a \right) \hat{y} + R \hat{\rho}
\end{array}\nonumber \\ & \notag  \hspace{4cm} \ \ \ \textrm{with} \ \ \  j=0, n-1. \\
\boldsymbol{R}^B_{l,j}=& 
\Bigg\{ \begin{array}{c} 
 R \ \frac{\pi (2 j + 1)}{n} \hat{\varphi} + \left( l \lambda + \frac{1}{2 
\sqrt{3}} a \right) \hat{y} + R \hat{\rho} \\ 
 R \ \frac{2 \pi  j }{n} \hat{\varphi} + \left(  l \lambda + \frac{2}{
\sqrt{3}} a \right) \hat{y} + R \hat{\rho}
\end{array} \\ & \notag  \hspace{4cm}  \ \ \ \textrm{with} \ \ \  j=0, n-1.
\end{align}
The distance between two unit cells along the axial direction is $\lambda = \sqrt{3}  a $. 
We insert coordinates \eqref{zz-coor} into \eqref{eq.int-phi2}, obtaining
\begin{align}
\notag
 &\rho_{\alpha\alpha'} = \frac{1}{2n}  \sum_{\eta=A}^B (F^{\eta}_{\tau \alpha k})^{*} F^{\eta}_{\tau  \alpha' k+q}  \\ & \times\quad  \sum^{n-1}_{j=0} 
 \Big[  e^{-i R G_{\perp} \! \cos(\frac{2 \pi j}{n})}  +   \label{eq.Jac-Ang-zz} e^{-i R G_{\perp} \! \cos(\frac{\pi (2j+1)}{n})}  \Big].
\end{align}
We rewrite the two exponentials as a sum of Bessel functions by using the Jacobi-Anger identity: 
\begin{align}
\label{eq.Jac-Ang}
e^{i x \cos \zeta} =J_0(x) + 2\sum_{m=1}^{\infty} i^m J_m(x) \cos(m\zeta).
\end{align}
Then \eqref{eq.Jac-Ang-zz} becomes:
\begin{align} \notag
 &  \rho_{\alpha\alpha'} = \frac{1}{2n}  \sum_{\eta=A}^B (F^{\eta}_{\tau \alpha k})^{*} F^{\eta}_{\tau  \alpha' k+q}   \sum^{n-1}_{j=0} \Big\{ 2 J_0(x)
 \\&+\quad 2\sum_{m=1}^{\infty} \Bigg[i^m J_m(x)  \cos \left( \! \frac{\pi m j}{n} \! \right)  \\&  + \quad  i^m J_m(x) \cos \! \left( \! \frac{\pi m (2j + 1) }{n} \right) \Bigg]\Big\}.
\end{align}
We now sum over $j$. The first term is obvious. In the second one, we exchange the order of the sums over $j$ and $m$, and use the
identities
\begin{multline}
\sum^{n-1}_{j=0} \! i^m J_m(x) \cos \! \left( \! \frac{\pi m j}{n} \! \right) = \\
\Bigg\{ \! \! \begin{array}{cc}
  n  \ i^m J_m(x) & \  \textrm{if} \  m \  \textrm{is a multiple of } n\\
  0 & \ \ \textrm{otherwise}
\end{array},
\end{multline}
\begin{multline}
\sum^{n-1}_{j=0} \! i^m J_m(x) \cos \! \left( \! \frac{\pi m (2j + 1) }{n} \! \right)  = \\ 
\Bigg\{ \! \! \begin{array}{cc}
  (-1)^{\frac{m}{n}} n \ i^m  J_m(x) & \  \textrm{if} \ m \ \textrm{is a multiple of }n   \\ 
  0  & \ \ \ \ \textrm{otherwise}  
\end{array}.
\end{multline}
Putting everything together, the overlap integral becomes:
\begin{multline}
\langle \alpha k|  e^{-i (\boldsymbol{G} + \boldsymbol{q}) \cdot \boldsymbol{r}} | \alpha' k + q  \rangle = \\ \left( (F^{ A}_{\tau \alpha k})^{*} F^{A}_{\tau \alpha' k+q} + (F^{ B}_{\tau \alpha k})^{*} F^{ B}_{\tau \alpha' k+q} \right)  \\ \left[  J_0(R G_{\perp}) + 2 \sum_{m=1}^{\infty} (-1)^m  J_{2mn} (R G_{\perp}) \right].
\end{multline}
As the Bessel functions of large multiples of $n$ decay very rapidly, we only consider the first term in the sum over $m$. The intraband and interband overlap integrals are 
\begin{multline}
 \langle c k|  e^{-i (\boldsymbol{G} + \boldsymbol{q}) \cdot \boldsymbol{r}} | c k+q  \rangle = \\ \frac{1}{2}  \left( \frac{k (k+q) +k_{\tau}^2-i q k_{\tau}}{\sqrt{k^2+k_{\tau}^2}\sqrt{(k+q)^2+k_{\tau}^2}} + 1 \right)\quad \times \\ (J_0(R G_{\perp}) + 2 (-1)^n  J_{2n} (R G_{\perp})) 
\end{multline}
and
 \begin{multline}
\langle c k|  e^{-i (\boldsymbol{G} + \boldsymbol{q}) \cdot \boldsymbol{r}} | v k+q  \rangle = \\  \\\frac{1}{2} \left( \frac{k (k+q) +k_{\tau}^2-i q k_{\tau}}{\sqrt{k^2+k_{\tau}^2}\sqrt{(k+q)^2+k_{\tau}^2}} - 1 \right) \quad \times \\ (J_0(R G_{\perp}) + 2 (-1)^n  J_{2n} (R G_{\perp})),
\end{multline}
respectively.

\section{Overlap integrals of armchair carbon nanotubes}
\label{armchair-ex-ov}

Armchair nanotubes $(n,n)$ may be thought of as built by rolling the graphene sheet along the chiral vector $\boldsymbol{\mathcal{C}} = 2 n \boldsymbol{a} + n \boldsymbol{b}$, which covers $2n$ units cell of graphene. There are $n$ atoms per sublattice, with positions:
\begin{eqnarray}
\begin{aligned}
\label{ac-coor}
\boldsymbol{R}^A_{l,j}=  R \ \frac{\pi j}{n} \hat{\varphi}  + \left(l \lambda  -  (-1)^j \frac{a}{4} \right)  \hat{y} & + R \hat{\rho} \\  & \ \ \ \textrm{with} \ \ \ j =0, 2n-1, \\
\boldsymbol{R}^B_{l,j}=  R \ \left( \frac{\pi j}{n} \hat{\varphi} + \frac{\pi}{3n} \right) \hat{\varphi}  +  \Big(l \lambda  - &
 (-1)^j \frac{a}{4} \Big) \hat{y} + R \hat{\rho}\\ & \ \ \ \textrm{with} \ \ \  j=0, 2n-1.
\end{aligned}
\end{eqnarray}
The distance between two subsequent unit cells along the axial direction is $\lambda=a$. To compute the overlap integrals $\rho_{\alpha\alpha'}$, we follow the same procedure used for zigzag tubes in the previous appendix. 
After integrating over $\varphi$, we obtain: 
\begin{align} \notag
& \rho_{\alpha\alpha'} = \frac{1}{2n} \sum^{2n-1}_{j=0} \Big[ (F^{ A}_{\tau \alpha k})^{*}  \notag  F^{A}_{\tau \alpha' k+q} e^{-i R G_{\perp} \cos \left(\frac{\pi j}{n} \right)}\\& + \quad (F^{B}_{\tau \alpha k})^{*} F^{B}_{\tau \alpha' k+q}  e^{-i R G_{\perp} \cos \left(\frac{\pi j}{n} + \frac{\pi}{3 n} \right) } \Big].
\end{align}
Again, using the Jacobi-Anger identity \eqref{eq.Jac-Ang}, we rewrite the exponentials as a sum of Bessel functions of different orders, $J_m$. We then obtain two sums over indexes $j$ and $m$,  and evaluate the sums over $j$ for given $m$, according to
\begin{multline}
\sum^{2n-1}_{j=0} \!  i^m J_m(x) \cos \! \left( \! \frac{\pi m j }{n} \! \right) \! = \\ \!
\Bigg\{ \! \! \begin{array}{cc}
      2n  \ i^m J_m(x) & \ \textrm{if} \ m \ \textrm{is a multiple of } 2n\\
     0 & \ \textrm{otherwise}
\end{array}
\end{multline}
and
\begin{multline}
\sum^{2n-1}_{j=0} \!  i^m J_m(x) \cos \! \left( \! \frac{\pi m (3 j + 1) }{3 n} \! \right) \! = \\ \! \Bigg\{ \! \! \begin{array}{cc}
   2 n  \cos \! \left( \! \frac{m \pi}{3} \! \right) \ i^m  J_m(x) & \ \textrm{if} \ m \ \textrm{is a multiple of }2n   \\
     0  & \ \textrm{otherwise}  
\end{array}.
\end{multline}
The resulting overlap integral is:
\begin{eqnarray}
\begin{aligned}
& \langle \alpha k|  e^{-i (\boldsymbol{G} + \boldsymbol{q}) \cdot \boldsymbol{r}} | \alpha' k+q  \rangle  = \\ & \frac{1}{2} \Bigg\{ (F^{A}_{\tau \alpha k})^{*} F^{ A}_{\tau \alpha' k+q} \! \left[ \! J_0(R G_{\perp}) + 2 \! \sum_{m=1}^{\infty} (-1)^m  J_{2mn} (R G_{\perp}) \! \right]     \\ & +  (F^{ B}_{\tau \alpha k})^{*} F^{B}_{\tau \alpha' k+q} \! \Bigg[ \! J_0(R G_{\perp}) + 2 \! \sum_{m=1}^{\infty} \cos \! \left( \! \frac{2 m\pi}{3} \! \right)\quad\times \\ & (-1)^m  J_{2mn} (R G_{\perp}) \! \Bigg]  \Bigg\}.
\end{aligned}
\end{eqnarray}
As before, it is sufficient to retain the first addendum of the sum over $m$. The intraband and interband overlap integrals of armchair nanotubes are
\begin{eqnarray}
\begin{aligned}
& \langle c k|  e^{-i (\boldsymbol{G} + \boldsymbol{q}) \cdot \boldsymbol{r}} | c k+q  \rangle  = \\ &\frac{1}{2} \left[ \frac{k(k+q)+k_{\tau}^2 - i q k_{\tau}}{\sqrt{k_{\tau}^2+k^2}\sqrt{k_{\tau}^2+(k+q)^2}} + 1  \right] J_0(R G_{\perp})\quad + \\  & \frac{(-1)^n}{2} \left[ 2 \frac{k(k+q)+k_{\tau}^2 - i q k_{\tau}} {\sqrt{k_{\tau}^2+k^2}\sqrt{k_{\tau}^2+(k+q)^2}} - 1  \right] J_{2n}(R G_{\perp})
\end{aligned}
\end{eqnarray}
and
\begin{eqnarray}
\begin{aligned}
& \langle c k|  e^{-i (\boldsymbol{G} + \boldsymbol{q}) \cdot \boldsymbol{r}} | v k+q  \rangle  = \\& \frac{1}{2} \left[ \frac{k(k+q)+k_{\tau}^2 - i q k_{\tau}}{\sqrt{k_{\tau}^2+k^2}\sqrt{k_{\tau}^2+(k+q)^2}} - 1  \right] J_0(R G_{\perp}) \quad + \\ & \frac{(-1)^n}{2} \! \left[ 2 \frac{k(k+q)+k_{\tau}^2 - i q k_{\tau}} {\sqrt{k_{\tau}^2+k^2}\sqrt{k_{\tau}^2+(k+q)^2}} + 1  \right] \! J_{2n}(R G_{\perp}),
\end{aligned}
\end{eqnarray}
respectively.

\section{Limiting form
of the Coulomb potential at long wavelength}
\label{Coul-analysis}

In the main text, we have expanded the Coulomb potential through two different Fourier decompositions. The first one is the Fourier transform \eqref{eq:Vcyl} of the Coulomb  potential on  a  uniform  cylindrical surface, $V_{\text{cyl}}(m,q)$. The second one is the three-dimensional Fourier transform \eqref{Coulomb} of the truncated Coulomb potential, $v(\mathbf{q} + \mathbf{G})$. Independently from the Fourier decomposition of choice,   the long-range, macroscopic behaviour of $v(\mathbf{q} + \mathbf{G})$  and $V_{\text{cyl}}(m,q)$
in the limit $q\rightarrow 0$ must be the same, as both forms derive from the same real-space potential, $e^2/r$ ($r$ is the radial distance in spherical coordinates). To show this, we constrain the Fourier transformation to a finite nanotube length, $A$ (we use the symbols $\Tilde{v}$ and $\Tilde{V}_{\text{cyl}}$ to identify the quantities obtained in this way). For $\Tilde{v}$ one has:
\begin{align}
\Tilde{v}(q,\mathbf{G_{\perp}})= \frac{e^2}{A \pi \mathfrak{R}^2} \int_{-A/2}^{A/2}  \! \! \! \! \! \! \! \! d y \int_{0}^{2 \pi} \! \! \! \! \! \! d\varphi \int_{0}^{\mathfrak{R}} \! \! \! \! d\varrho  \frac{\varrho}{\sqrt{\varrho^2+y^2}} e^{i q y} e^{i \varrho G_{\perp} \cos\varphi}.
\end{align}
We now take the limit $q \rightarrow 0$:
\begin{align}
\Tilde{v}(0,\mathbf{G_{\perp}})= \frac{e^2}{A \pi \mathfrak{R}^2}\int_{-A/2}^{A/2}  \! \! \! \! \! \! \! \! d y \int_{0}^{2 \pi} \! \! \! \! \! \! d\varphi \int_{0}^{\mathfrak{R}} \! \! \! \! d\varrho \frac{\varrho}{\sqrt{\varrho^2+y^2}}  e^{i \varrho G_{\perp} \cos\varphi},
\end{align}
which gives
\begin{multline}
\Tilde{v}(0,\mathbf{G_{\perp}})= \frac{4 e^2}{A (G_{\perp} \mathfrak{R})^2}\Big[1 - J_0( G_{\perp} \mathfrak{R}) + G_{\perp} \mathfrak{R} \\ J_1( G_{\perp} \mathfrak{R}) \log\left(\frac{A}{\mathfrak{R}} \right) \Big].  
\end{multline}
The long-wavelength, macroscopic limit is:
\begin{eqnarray}
\label{eq.reg-div1}
\Tilde{v}(0,0)= \frac{e^2}{A}\left[1 + 2 \log\left(\frac{A}{\mathfrak{R}} \right)\right].
\end{eqnarray}
For $\Tilde{V}_{\text{cyl}}$ one has:
\begin{align}
\Tilde{V}_{\text{cyl}}(m,q) = & \frac{e^2}{2 A \pi R} \int_{-A/2}^{A/2} \! \! \! \! \! \! \! \! d y \int_{0}^{2 \pi} \! \! \! \! \! \! d\varphi   \frac{e^{ i m \varphi} \ e^{i q y}}{\sqrt{4 R^2 \sin^2\left(\frac{\varphi}{2}\right) + y^2}},
\end{align}
and in the limit $q \rightarrow 0$:
\begin{align}
\Tilde{V}_{\text{cyl}}(m,0) = & \frac{e^2}{2 A \pi R} \int_{-A/2}^{A/2} \! \! \! \! \! \! \! \! d y \int_{0}^{2 \pi} \! \! \! \! \! \! d\varphi  \frac{e^{ i m \varphi}}{\sqrt{4 R^2 \sin^2\left(\frac{\varphi}{2 }\right) + y^2}},
\end{align}
that is
\begin{eqnarray}
\Tilde{V}_{\text{cyl}}(m,0) = \left\{ \begin{array}{cc}
     \frac{e^2}{A m}   & \textrm{if} \ \ \ m \neq 0 \\
     \frac{2 e^2}{A}  \log \left(\frac{A}{R} \right) & \textrm{if} \ \ \ m = 0
\end{array} \right. .
\end{eqnarray}
The macroscopic term amounts to:
\begin{eqnarray}
\label{eq.reg-div2}
\Tilde{V}_{\text{cyl}}(0,0) =  \frac{2 e^2}{A}  \log \left(\frac{A}{R} \right).
\end{eqnarray}
We recall that the nanotube length $A$ is linked to the sampling of the Brillouin zone, $A={2 \pi}/\text{d} q$. As a consequence both macroscopic potentials \eqref{eq.reg-div1} and \eqref{eq.reg-div2} exhibit an analogous logarithmic divergence of the kind  $\sim - \log(R \,\text{d} q)$, leading to the same long-range behaviour.


\end{document}